%% file: yrast.tex
\documentclass[reviewcopy]{elsarticle}
\usepackage{longtable}
\usepackage{natbib,mathptmx}
\usepackage{amsmath,amssymb}
\usepackage{graphicx,array}
\usepackage{adndt}
\usepackage{color}

\biboptions{square,sort&compress}
\bibpunct[]{[}{]}{,}{n}{}{;}
\begin{document}
\begin{frontmatter}
\journal{Atomic Data and Nuclear Data Tables}
\title{Empirical predictions of yrast energies in even-even nuclei}
\author[china]{Guanghao Jin}
\author[inha]{Dooyoung Kim}
\author[inha]{Byeongnoh Kim}
\author[inha]{Jin-Hee Yoon}
\author[inha]{Dongwoo Cha\corref{cor1}}
\ead{E-mail: dcha@inha.ac.kr}
\cortext[cor1]{Corresponding author.}
\address[china]{Department of Fundamental Subject, Tianjin Institute of Urban Construction,
  Tianjin 300384, People's Republic of China}
\address[inha]{Department of Physics, Inha University, Incheon 402-751, South Korea}

\date{01.07.2010} 

\begin{abstract}
  The lowest excitation energies of the given multipole $J^\pi$ state (the $J^\pi$ yrast energies) are given for even-even nuclei throughout the entire periodic table. The yrast energies were calculated using the recently proposed empirical formula that depends only on the mass number $A$, and the valence nucleon numbers $N_p$ and $N_n$. We provide a complete tabulation and plots of the yrast energies calculated using the empirical formula together with the ones measured for the natural parity states up to $10^+$ and for the unnatural parity states up to $10^-$ with the hope of encouraging active study on the possible origin of the relationship between the yrast energies, as revealed by the empirical formula.
\end{abstract}

\end{frontmatter}

\newpage

\tableofcontents

\listofDtables
\listofDfigures

\vskip3pc


\section{Introduction}

The exhaustive compilation of experimental results for the reduced electric quadrupole transition probability, $B(E2)$, between the $0^+$ ground state and first $2^+$ state in even-even nuclei by Raman {\it et al.} provided a rare opportunity to perform a systematic study of the relevant nuclear properties throughout the periodic table \cite{Raman}. They have even suggested the `global best fit', $B(E2) = (2.57 \pm 0.45) E^{-1} Z^2 A^{2/3}$, to the measured $B(E2)$ values in terms of simply the atomic number $Z$ and mass number $A$. However, knowledge of the measured excitation energy $E$ of the first $2^+$ state was still required to make acceptable global predictions in terms of just $Z$ and $A$.

Meanwhile, the valence nucleon numbers $N_p$ and $N_n$ have been used widely to parameterize various nuclear properties phenomenologically. Hamamoto was the first to recognize the utility of the valence nucleon numbers and show that the square root of the ratio of the measured and the single-particle $B(E2)$ values, $[B(E2)_{\rm exp} / B(E2)_{\rm sp}]^{1/2}$, is roughly proportional to the product $N_p N_n$ \cite{Hamamoto}. In addition, Casten noticed that a simple pattern appeared when nuclear data concerning nuclear deformation was plotted as a function of the product $N_p N_n$ and such phenomenon was referred to as the $N_p N_n$ scheme. Indeed, the $N_p N_n$ scheme has been used extensively and successfully for more than two decades to correlate large volumes of data on the collective degrees of freedom in nuclei \cite{Casten1,Casten2}.

Recently, we reported the empirical findings of a simple formula that could reproduce the first $2^+$ excitation energy (the $2^+$ yrast energy) in even-even nuclei \cite{Ha1}. The idea for the empirical formula was first envisaged by inspecting Figs. I(a), II(a), and III(a) of Ref.\,\cite{Raman} where the measured $2^+$ yrast energy in 557 even-even nuclei was plotted as function of the mass number $A$, the atomic number $Z$, and the neutron number $N$, respectively. Later, the same empirical formula was shown to be capable of describing the main trends of the yrast energies of not only the electric quadrupole states but also the natural parity even as well as odd multipole states up to $10^+$ found in all even-even nuclei \cite{Kim1,Jin1}. Furthermore, the yrast energies followed the same empirical formula even for unnatural parity states \cite{Kim2}.

Thus it is now evident that this empirical formula characterizes the overall shape of the yrast energies for all multipoles, including both natural and unnatural parity states. Once it has been established that there is a universal relationship between the yrast energies, such as our empirical formula, it is natural to imagine that there would be some underlying dynamical origin for such a relationship. Unfortunately, the origin is unclear. This paper provides a complete tabulation and plots of the yrast energies calculated using the empirical formula together with the measured ones for the natural parity states up to $10^+$ and also for the unnatural parity states up to $10^-$ with the hope of soliciting an active study on the possible origins of the relationship between yrast energies.

\section{Empirical formula and yrast energy distribution}

\begin{figure}[ht!]
\begin{center}
\includegraphics[width=.5\linewidth]{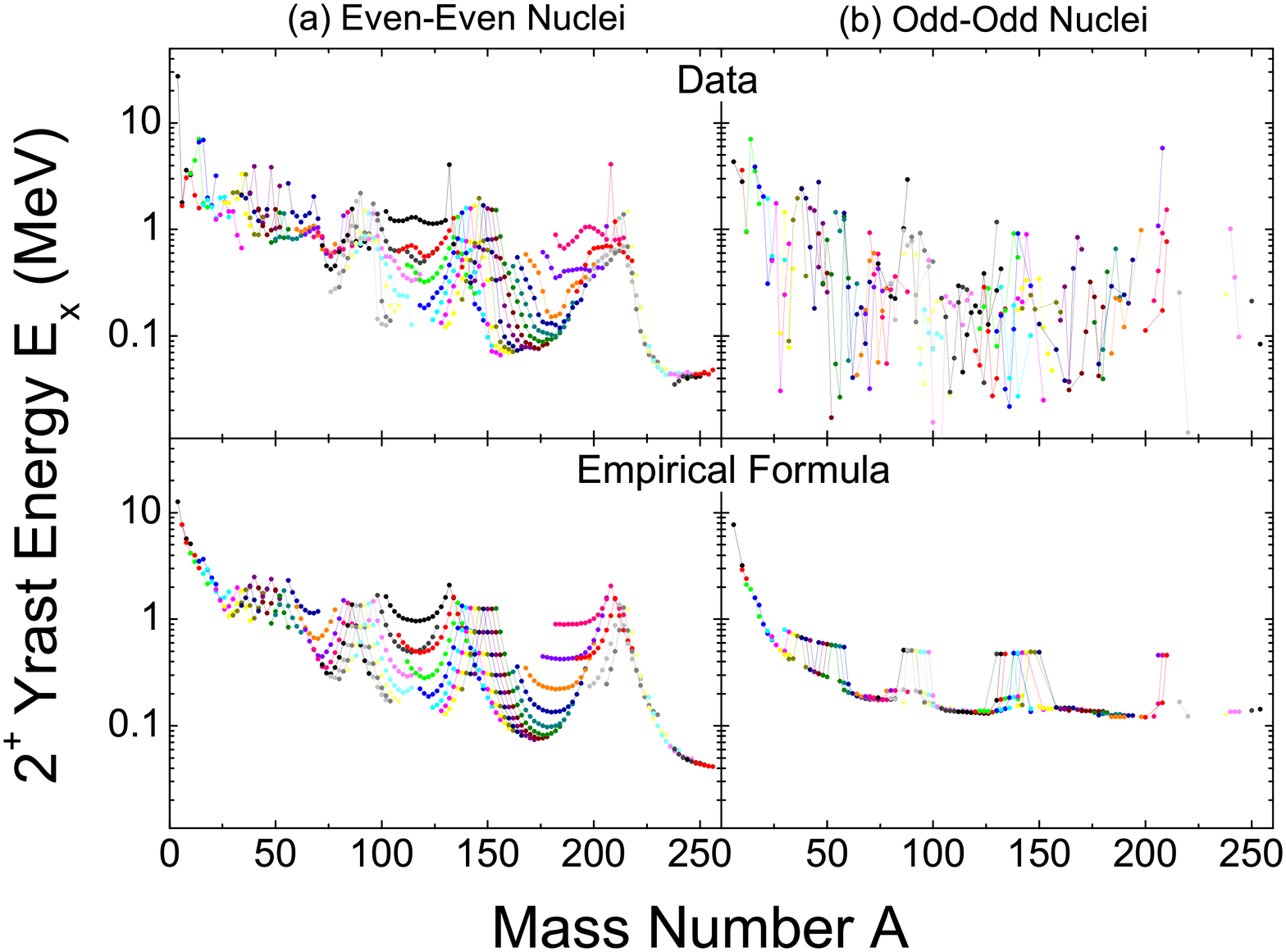}
\caption {(Color online) Yrast energies of $2^+$ states in (a) even-even (left panels) and (b) odd-odd nuclei (right panels). The data points are connected along the isotopic chains. The upper two panels show the measured $2^+$ yrast energies while the lower two panels show those calculated by using the empirical formula. The measured $2^+$ yrast energies are quoted from Ref.\,\cite{Raman} for even-even nuclei and collected from the ENSDF database for odd-odd nuclei \cite{ENSDF}.}
\label{fig-1}
\end{center}
\end{figure}

The empirical formula, mentioned in the previous section, was first introduced to find a simple formula that could reproduce the graph of the $2^+$ yrast energies shown in the upper panel of Fig.\,\ref{fig-1}(a) which shows the data quoted from Ref.\,\cite{Raman}, where the best known values of the $2^+$ yrast energies were compiled for even-even nuclei. However, one does not normally attempt to describe any graph with many spikes, such as the one shown in Fig.\,\ref{fig-1}(a), using a formula that depends on only smoothly changing variables. Even after adopting the valence nucleon numbers $N_p$ and $N_n$ as well as the mass number $A$, it was barely possible to devise an empirical formula that could describe complicated graphs of the $2^+$ yrast energies measured from all even-even nuclei throughout the entire periodic table. The valence proton (neutron) number $N_p$ ($N_n$) of a nucleus with an atomic (neutron) number $Z$ ($N$) is defined as
\begin{equation} \label{Val}
N_p (N_n) = \left\{ \begin{array}{ll}
   Z (N)-N_{c-1} \, \, \mbox{if $N_{c-1} < Z (N) \leq  M_c$}\\
   N_c - Z (N) \, \, \, \mbox{if $M_c < Z (N) \leq N_c$} \, ,\end{array} \right.
\end{equation}
where $N_c$ is the magic number for the $c$-th major shell, and $M_c$ is the average of the two adjacent magic numbers, $(N_{c-1}+N_c)/2$, which corresponds to the number of nucleons contained in the mid-shell nucleus of the $c$-th major shell. The valence nucleon numbers $N_p$ and $N_n$ repeat the positive integer numbers from zero whenever the atomic number $Z$ or the neutron number $N$ crosses one of the major shell boundaries.

The original form of the empirical formula first introduced in Ref.\,\cite{Ha1} for the $2^+$ yrast energy in even-even nuclei was written as
\begin{equation} \label{E4}
E_x = \alpha A^{-\gamma} + \beta \left[ e^{- \lambda   N_p } + e^{- \lambda   N_n } \right],
\end{equation}
where $\alpha$, $\gamma$, $\beta$, and $\lambda$ are four model parameters to be fitted from the data. However, after testing different formulae with several other forms, including a term with the product $N_p N_n$, the following six-parameter form was chosen as the best expression for the yrast energy $E_x$ in even-even nuclei \cite{Jin2}:
\begin{equation} \label{E6}
E_x  = \alpha  A^{-\gamma}  + \beta_p  e^{- \lambda_p   N_p }  + \beta_n e^{- \lambda_n N_n }.
\end{equation}
Here, the parameters $\beta$ and $\lambda$ in Eq.\,(\ref{E4}) are split into $\beta_p$, $\beta_n$ and $\lambda_p$, $\lambda_n$, respectively. This considers the fact that protons and neutrons make different contributions to the yrast energy $E_x$.

The $2^+$ yrast energies calculated from Eq.\,(\ref{E6}) are plotted in the lower panel of Fig.\,\ref{fig-1}(b). We can find a very close similarity between the curves in the upper and lower panels in Fig.\,\ref{fig-1}(a), the data and our calculated results, respectively. (In the electronic version, the color code for an isotopic chain in the upper panel is the same as the color code for the corresponding isotopic chain in the lower panel.) Although it is remarkable that a simple formula, such as Eq.\,(\ref{E6}), can reproduce the data both qualitatively and quantitatively to some extent, it is better to discuss what it means to claim that there is a certain meaningful relationship between a myriad of data points. This issue is raised because our empirical formula was sometimes critiqued for its use of too many free parameters. However, as a counter example, where no simple relationship can be easily found, consider the same $2^+$ yrast energies but measured in odd-odd nuclei, which are plotted in the upper panel of Fig.\,\ref{fig-1}(b). The measured $2^+$ yrast energies in odd-odd nuclei were collected from the ENSDF database \cite{ENSDF}. The same $2^+$ yrast energies, in odd-odd nuclei, calculated using the empirical formula are shown in the lower panel of Fig.\,\ref{fig-1}(b). In contrast to the $2^+$ yrast energies in even-even nuclei, a comparison of the graphs shown in the upper and lower panels of Fig.\,\ref{fig-1}(b) suggests that the empirical formula can never represent the $2^+$ yrast energy data measured from odd-odd nuclei.

\begin{figure}[ht!]
\begin{center}
\includegraphics[width=.5\linewidth]{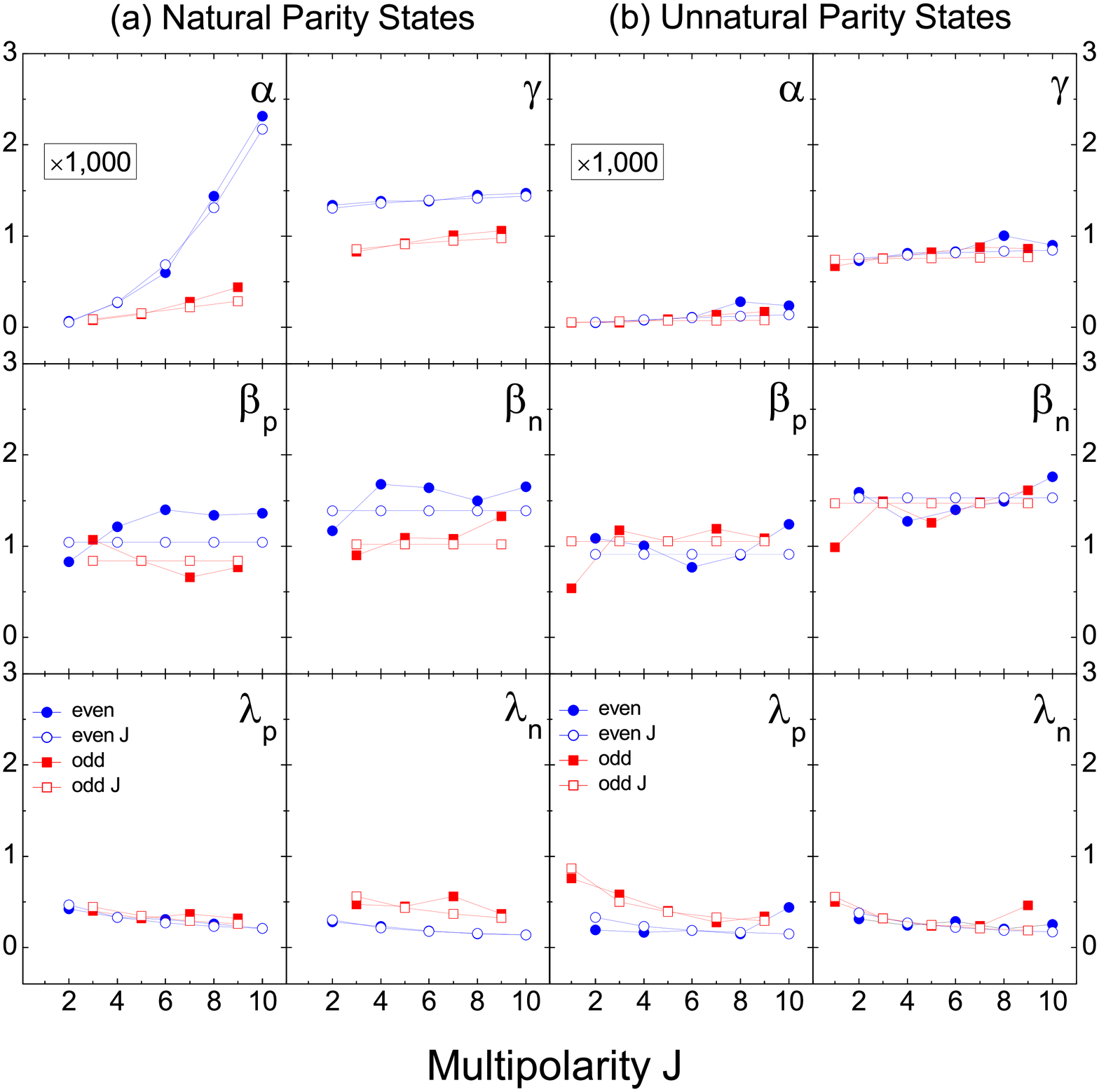}
\caption {(Color online) Six parameters that appear in Eq.\,(\ref{E6}). Solid symbols denote the parameter values determined for each multipoles and open symbols reflect the results obtained by using the spin dependent empirical formula. Circles are for the even multipole states while squares are for the odd multipole states. The parameter values for the natural parity states are quoted from Refs.\,\cite{Kim1} and \cite{Jin2} and those for the unnatural parity states are quoted from Ref.\,\cite{Kim2}.}
\label{fig-2}
\end{center}
\end{figure}

The six model parameters $\alpha$, $\gamma$, $\beta_p$, $\beta_n$, $\lambda_p$, and $\lambda_n$ of Eq.\,(\ref{E6}) can be determined easily and unambiguously using the usual least-squares-fitting procedure. The values adopted for these six parameters are listed in Table\,\ref{tab-a} together with the total data points $N_0$ for each multipole state. The values are quoted from Refs.\,\cite{Kim1} and \cite{Jin1} for the natural parity states and from \cite{Kim2} for the unnatural parity states. The values are also shown in Fig.\,\ref{fig-2} with solid circles (even multipoles) and solid squares (odd multipoles).

The parameter values listed in Table\,\ref{tab-a} were fitted to each multipole separately. However, one can devise a formula that can be used for different multipoles with the same spin dependent parameter values after replacing the four parameters $\alpha$, $\gamma$, $\lambda_p$, and $\lambda_n$ with
\begin{equation} \label{new_p}
\alpha = \alpha_0 J^a ,~~\gamma = \gamma_0 J^c , ~~  \lambda_p={\lambda_p^0  \over \sqrt{J}} ~~~{\rm and} ~~~ \lambda_n= {\lambda_n^0 \over \sqrt{J}},
\end{equation}
where $a$ and $c$ are additional parameters introduced to give the proper $J$ dependence of $\alpha$ and $\gamma$, respectively, and $\lambda_p^0$ and $\lambda_n^0$ are new $J$-independent parameters that were fitted in place of $\lambda_p$ and $\lambda_n$, respectively \cite{Jin3}. The spin-dependent empirical formula can now be expressed as follows:
\begin{equation} \label{sE}
E_x = \alpha_0 J^a A^{-\gamma_0 J^c} + \beta_p  e^{- {\lambda_p^0 N_p \over \sqrt{J}}} + \beta_n e^{- {\lambda_n^0 N_n \over \sqrt{J}}}.
\end{equation}
Now, the eight parameters $\alpha_0$, $a$, $\gamma_0$, $c$, $\beta_p$, $\beta_n$, $\lambda_p^0$, and $\lambda_n^0$ are determined using the yrast energies of all the even or odd multipoles of the natural or unnatural parity states.

The results for the eight parameters in Eq.\,(\ref{sE}), which are quoted from Ref.\,\cite{Jin3} (for natural parity states) and Ref.\,\cite{Kim3} (for unnatural parity states), are listed in Table\,\ref{tab-b}, together with the number of total data points $N_0$, which were included in the fitting procedure. The original six parameters estimated using the spin dependent formula, Eq.\,(\ref{sE}), are shown in Fig.\,\ref{fig-2} with open circles (even multipoles) and open squares (odd multipoles). By comparing the open symbols with the corresponding solid ones in Fig.\,\ref{fig-2}, it is evident that the agreement between the parameters obtained by fitting each multipole separately and those parameters obtained using the spin dependent formula is quite impressive. It was also pointed out that the increase in the $\chi^2$ value after using the spin dependent empirical formula amounts to only $\sim 5 \%$ \cite{Jin3}. Therefore, the spin dependent empirical formula, Eq.\,(\ref{sE}), reproduces the results of the spin independent case, Eq.\,(\ref{E6}), with fewer parameters.

From the results for the six parameters shown in Fig.\,\ref{fig-2} and Tables\,\ref{tab-a} and Table\,\ref{tab-b}, we can make the following observations on the yrast energy distributions. The parameters $\alpha$ and $\gamma$, which belong to the mass-dependent term of Eq.\,(\ref{E6}), show a different characteristic dependence on $J$ according to whether they represent even or odd multipole states of the natural or unnatural parity states. In particular, Table\,\ref{tab-b} shows that $\alpha$ has an almost quadratic dependence on $J$ for even multipoles and an almost linear dependence for odd multipoles in the case of the natural parity states. However, in the case of the unnatural parity states, $\alpha$ becomes practically constant over $J$ for both even and odd multipoles.

\begin{figure}[ht!]
\begin{center}
\includegraphics[width=.6\linewidth]{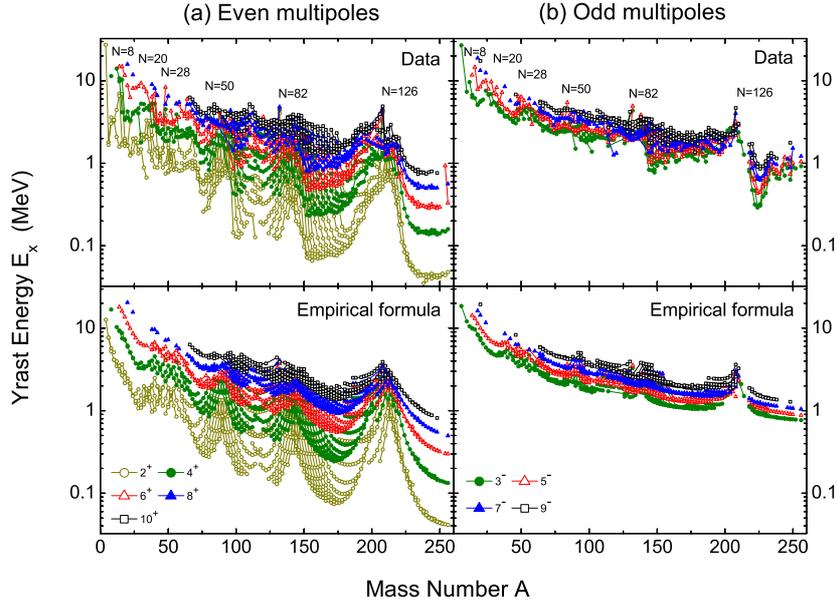}
\caption {(Color online) Yrast energies of the natural parity (a) even multipole and (b) odd multipole states in even-even nuclei. The upper two panels show the measured yrast energies while the lower two panels show those calculated by using the empirical formula given by Eq.\,(\ref{E6}). The measured excitation energies were collected from the ENSDF database \cite{ENSDF}.}
\label{fig-3}
\end{center}
\end{figure}

\begin{figure}[ht!]
\begin{center}
\includegraphics[width=.55\linewidth]{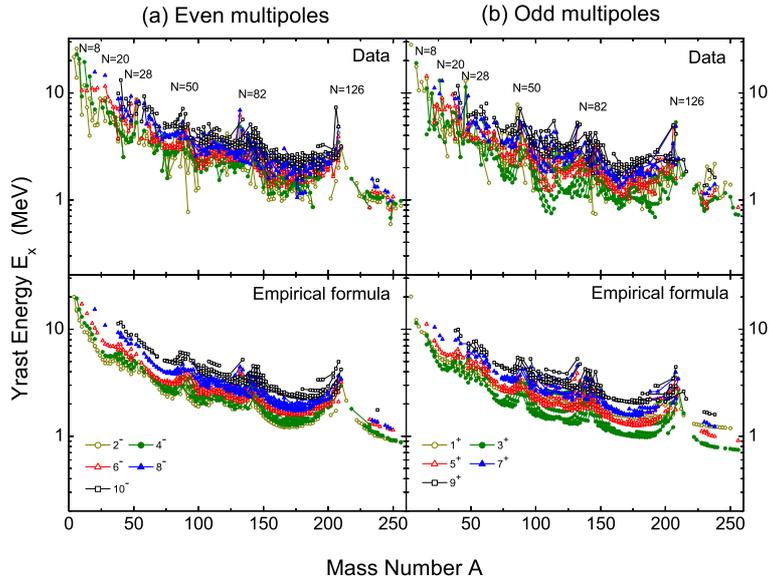}
\caption {(Color online) Same as Fig.\,\ref{fig-3}, but for the unnatural parity states. The measured excitation energies were collected from the ENSDF database \cite{ENSDF}.}
\label{fig-4}
\end{center}
\end{figure}

Of the six parameters, the first two parameters, $\alpha$ and $\gamma$, determine the gross behavior of the yrast energy distributions. Therefore, the difference between the values of $\alpha$ and $\gamma$ determines the sharp distinction in the gross shape of the yrast energy distributions shown in Figs.\,\ref{fig-3} and \ref{fig-4} between the even and odd multipoles of the natural or unnatural states. For the natural parity states, we show the measured (upper panel) and calculated (lower panel) yrast energies in even-even nuclei for the even multipole states including $2^+$, $4^+$, $6^+$, $8^+$, and $10^+$, in Fig.\,\ref{fig-3}(a), while we also show those for the odd multipole states, including $3^-$, $5^-$, $7^-$, and $9^-$ in Fig.\,\ref{fig-3}(b). The dipole ($1^-$) yrast energies are excluded from Fig.\,\ref{fig-3}(b) because they do not follow the common pattern of the other odd multipole cases shown in Fig.\,\ref{fig-3}(b) \cite{Jin1}. Similarly, for the unnatural parity states, we show the yrast energies in even-even nuclei for the even multipole states including  $2^-$, $4^-$, $6^-$, $8^-$, and $10^-$, in Fig.\,\ref{fig-4}(a), whereas we show also those for the odd multipole states, including $1^+$, $3^+$, $5^+$, $7^+$, and $9^+$ in Fig.\,\ref{fig-4}(b).

By comparing the graphs shown in Fig.\,\ref{fig-3}(a) and (b), we immediately find for the natural parity states that although the yrast energies are getting larger as the multipole of the state increases for both the even and odd multipole states, the yrast energies of odd multipole states lie significantly closer together than those of the even multipole states. For the unnatural states, however, by comparing the graphs shown in Fig.\,\ref{fig-4}(a) and (b), we find that the overall shapes of the yrast energy distributions of the even and odd multipole states are quite similar. In contrast to the first two parameters, $\alpha$ and $\gamma$, the remaining four parameters $\beta_p$, $\beta_n$, $\lambda_p$, and $\lambda_n$, which belong to the terms involving the valence nucleon numbers, depend similarly on $J$ regardless of the multipole-parity relations of the states they belong to and determine the detailed shape of the yrast energy distributions within each major shell.

\bigskip

\section*{Acknowledgments}
This research was supported by Basic Science Research Program through the National Research Foundation of Korea(NRF) funded by the Ministry of Education, Science and Technology (2009-0075609).


\clearpage

\section*{Figures}

\clearpage
\renewcommand{\baselinestretch}{1.0}
\section*{Tables}
\begin{table}[ht!]
\caption{The values adopted for the six parameters in the empirical formula, Eq.\,(\ref{E6}), for the yrast energies $E_x$ of the natural parity states including even multipoles ($2^+$, $4^+$, $6^+$, $8^+$, and $10^+$) and odd multipoles ($1^-$, $3^-$, $5^-$, $7^-$, and $9^-$) and also of the unnatural parity states including even multipoles ($2^-$, $4^-$, $6^-$, $8^-$, and $10^-$) and odd multipoles ($1^+$, $3^+$, $5^+$, $7^+$, $9^+$, and $11^+$). These values are quoted from Ref.\,\cite{Kim1} and \cite{Jin1} for the natural parity states and from Ref.\,\cite{Kim2} for the unnatural parity states. The last column denotes the total number $N_0$ of the data points for the corresponding multipole state.}
\begin{tabular}{cccccccc}
\hline\hline
~~$J^\pi$~~~&~~~$\alpha$~~~&~~~$\gamma$~~~&~~~$\beta_p$~~~&~~~$\beta_n$~~~
&~~~$\lambda_p$~~~&~~~$\lambda_n$~~~&~~~$N_0$~~~~\\
&(MeV)&&(MeV)&(MeV)&&&\\
\hline
$2^+$&68&1.34&0.83&1.17&0.42&0.28&557\\
$4^+$&268&1.38&1.21&1.68&0.33&0.23&430\\
$6^+$&598&1.38&1.40&1.64&0.31&0.18&375\\
$8^+$&1,439&1.45&1.34&1.50&0.26&0.15&309\\
$10^+$&2,317&1.47&1.36&1.65&0.21&0.14&265\\
\hline
$1^-$&75&0.83&2.18&2.33&0.57&0.44&196\\
$3^-$&77&0.83&1.07&0.90&0.40&0.47&317\\
$5^-$&144&0.92&0.84&1.09&0.32&0.45&352\\
$7^-$&283&1.01&0.66&1.08&0.37&0.56&315\\
$9^-$&442&1.06&0.77&1.33&0.32&0.37&267\\
\hline\hline
$2^-$&48&0.73&1.09&1.59&0.19&0.31&246\\
$4^-$&75&0.81&1.00&1.27&0.17&0.24&253\\
$6^-$&108&0.83&0.77&1.40&0.19&0.28&248\\
$8^-$&277&1.00&0.90&1.49&0.15&0.20&230\\
$10^-$&238&0.90&1.24&1.76&0.44&0.25&199\\
\hline
$1^+$&47&0.67&0.54&0.99&0.76&0.50&251\\
$3^+$&49&0.76&1.17&1.49&0.58&0.32&236\\
$5^+$&87&0.82&1.05&1.26&0.40&0.24&250\\
$7^+$&139&0.88&1.19&1.48&0.28&0.24&184\\
$9^+$&173&0.86&1.09&1.61&0.34&0.46&159\\
\hline \hline
\end{tabular}
\label{tab-a}
\end{table}

\begin{table}[ht!]
\caption{The values adopted for the eight parameters in the spin dependent empirical formula, Eq.\,(\ref{sE}), for the yrast energy of the natural parity even and odd multipole states (upper two rows) and of the unnatural parity even and odd multipole states (lower two rows). The values are quoted from Ref.\,\cite{Jin3} for the natural parity states and from Ref.\,\cite{Kim3} for the unnatural parity states. The last column denotes the total number $N_0$ of the data points included in the fitting procedure.}
\begin{tabular}{cccccccccc}
\hline\hline
~~$J^\pi$~~~&~~~$\alpha_0$~~~&~~~$a$~~~&~~~$\gamma_0$~~~&~~~$c$~~~&~~~$\beta_p$~~~&~~~$\beta_n$~~~
&~~~$\lambda_p^0$~~~&~~~$\lambda_n^0$~~~&~~~$N_0$~~~~\\
&(MeV)&&&&(MeV)&(MeV)&&&\\
\hline
Natural Even&11.94&2.26&1.25&0.06&1.04&1.39&0.66&0.43&1936\\
Natural Odd&28.15&1.05&0.75&0.12&0.84&1.02&0.77&0.97&1447\\
\hline
Unnatural Even&38.48&0.56&0.72&0.07&0.91&1.53&0.46&0.54&1176\\
Unnatural Odd&53.65&0.16&0.74&0.02&1.06&1.47&0.87&0.56&1197\\

\hline \hline
\end{tabular}
\label{tab-b}
\end{table}

\newpage

\TableExplanation
\bigskip

\section*{Table 1. ~~~~~~~~~Yrast energies of the natural parity even multipole states in even-even nuclei.}

\begin{tabular}{p{0.9cm}p{6in}}
&Yrast energies in MeV of the natural parity even multipole states including $J^\pi = 2^+$, $4^+$, $6^+$, $8^+$, and $10^+$ in even-even nuclei are given. The first number before the slash is the measured yrast energy and the second number after the slash is the calculated one by using Eq.\,(\ref{E6}).\\
\end{tabular}

\begin{tabular}{p{0.9cm}p{1in}p{5in}}
&Nuclide	& Even $Z$ and even $N$ nuclide studied \\
&$A$		& Mass number of the nuclide\\
&$N_p$      & Valence proton number of the nuclide\\
&$N_n$	    & Valence proton number of the nuclide\\
\end{tabular}

\bigskip

\section*{Table 2. ~~~~~~~~~Yrast energies of the natural parity odd multipole states in even-even nuclei.}

\begin{tabular}{p{0.9cm}p{6in}}
&Same as Table\,1 but for the natural parity odd multipole states including $J^\pi = 1^-$, $3^-$, $5^-$, $7^-$, and $9^-$.\\
\end{tabular}

\bigskip

\section*{Table 3. ~~~~~~~~~Yrast energies of the unnatural parity even multipole states in even-even nuclei.}

\begin{tabular}{p{0.9cm}p{6in}}
&Same as Table\,1 but for the unnatural parity even multipole states including $J^\pi = 2^-$, $4^-$, $6^-$, $8^-$, and $10^-$.\\
\end{tabular}

\bigskip

\section*{Table 4. ~~~~~~~~~Yrast energies of the unnatural parity odd multipole states in even-even nuclei.}

\begin{tabular}{p{0.9cm}p{6in}}
&Same as Table\,1 but for the unnatural parity odd multipole states including $J^\pi = 1^+$, $3^+$, $5^+$, $7^+$, and $9^+$.\\
\end{tabular}

\newpage

\GraphExplanation
\bigskip

\section*{Graph 1. ~~~~~~~${\bold 2^+}$ yrast energies in even-even nuclei.}

\begin{tabular}{p{0.9cm}p{6in}}
&(Color online) $2^+$ yrast energies in even-even nuclei are plotted against the mass number $A$ in the left two panels and against the product $N_pN_n$ between the valence proton number $N_p$ and the valence neutron number $N_n$ in the right two panels. The upper two panels show the measured yrast energies while the lower two panels show those calculated by the empirical formula, Eq.\,(\ref{E6}).\\
\end{tabular}

\bigskip
\section*{Graph 2. ~~~~~~~${\bold 4^+}$ yrast energies in even-even nuclei.}
\begin{tabular}{p{0.9cm}p{6in}}
&(Color online) Same as Graph 1 but for $4^+$ yrast energies.\\
\end{tabular}

\bigskip
\section*{Graph 3. ~~~~~~~${\bold 6^+}$ yrast energies in even-even nuclei.}
\begin{tabular}{p{0.9cm}p{6in}}
&(Color online) Same as Graph 1 but for $6^+$ yrast energies.\\
\end{tabular}

\bigskip
\section*{Graph 4. ~~~~~~~${\bold 8^+}$ yrast energies in even-even nuclei.}
\begin{tabular}{p{0.9cm}p{6in}}
&(Color online) Same as Graph 1 but for $8^+$ yrast energies.\\
\end{tabular}

\bigskip
\section*{Graph 5. ~~~~~~~${\bold 10^+}$ yrast energies in even-even nuclei.}
\begin{tabular}{p{0.9cm}p{6in}}
&(Color online) Same as Graph 1 but for $10^+$ yrast energies.\\
\end{tabular}

\bigskip
\section*{Graph 6. ~~~~~~~${\bold 1^-}$ yrast energies in even-even nuclei.}
\begin{tabular}{p{0.9cm}p{6in}}
&(Color online) Same as Graph 1 but for $1^-$ yrast energies.\\
\end{tabular}

\bigskip
\section*{Graph 7. ~~~~~~~${\bold 3^-}$ yrast energies in even-even nuclei.}
\begin{tabular}{p{0.9cm}p{6in}}
&(Color online) Same as Graph 1 but for $3^-$ yrast energies.\\
\end{tabular}

\bigskip
\section*{Graph 8. ~~~~~~~${\bold 5^-}$ yrast energies in even-even nuclei.}
\begin{tabular}{p{0.9cm}p{6in}}
&(Color online) Same as Graph 1 but for $5^-$ yrast energies.\\
\end{tabular}

\bigskip
\section*{Graph 9. ~~~~~~~${\bold 7^-}$ yrast energies in even-even nuclei.}
\begin{tabular}{p{0.9cm}p{6in}}
&(Color online) Same as Graph 1 but for $7^-$ yrast energies.\\
\end{tabular}

\bigskip
\section*{Graph 10. ~~~~~~~${\bold 9^-}$ yrast energies in even-even nuclei.}
\begin{tabular}{p{0.9cm}p{6in}}
&(Color online) Same as Graph 1 but for $9^-$ yrast energies.\\
\end{tabular}

\bigskip
\section*{Graph 11. ~~~~~~~${\bold 2^-}$ yrast energies in even-even nuclei.}
\begin{tabular}{p{0.9cm}p{6in}}
&(Color online) Same as Graph 1 but for $2^-$ yrast energies.\\
\end{tabular}

\bigskip
\section*{Graph 12. ~~~~~~~${\bold 4^-}$ yrast energies in even-even nuclei.}
\begin{tabular}{p{0.9cm}p{6in}}
&(Color online) Same as Graph 1 but for $4^-$ yrast energies.\\
\end{tabular}

\bigskip
\section*{Graph 13. ~~~~~~~${\bold 6^-}$ yrast energies in even-even nuclei.}
\begin{tabular}{p{0.9cm}p{6in}}
&(Color online) Same as Graph 1 but for $6^-$ yrast energies.\\
\end{tabular}

\bigskip
\section*{Graph 14. ~~~~~~~${\bold 8^-}$ yrast energies in even-even nuclei.}
\begin{tabular}{p{0.9cm}p{6in}}
&(Color online) Same as Graph 1 but for $8^-$ yrast energies.\\
\end{tabular}

\bigskip
\section*{Graph 15. ~~~~~~~${\bold 10^-}$ yrast energies in even-even nuclei.}
\begin{tabular}{p{0.9cm}p{6in}}
&(Color online) Same as Graph 1 but for $10^-$ yrast energies.\\
\end{tabular}

\bigskip
\section*{Graph 16. ~~~~~~~${\bold 1^+}$ yrast energies in even-even nuclei.}
\begin{tabular}{p{0.9cm}p{6in}}
&(Color online) Same as Graph 1 but for $1^+$ yrast energies.\\
\end{tabular}

\bigskip
\section*{Graph 17. ~~~~~~~${\bold 3^+}$ yrast energies in even-even nuclei.}
\begin{tabular}{p{0.9cm}p{6in}}
&(Color online) Same as Graph 1 but for $3^+$ yrast energies.\\
\end{tabular}

\bigskip
\section*{Graph 18. ~~~~~~~${\bold 5^+}$ yrast energies in even-even nuclei.}
\begin{tabular}{p{0.9cm}p{6in}}
&(Color online) Same as Graph 1 but for $5^+$ yrast energies.\\
\end{tabular}

\bigskip
\section*{Graph 19. ~~~~~~~${\bold 7^+}$ yrast energies in even-even nuclei.}
\begin{tabular}{p{0.9cm}p{6in}}
&(Color online) Same as Graph 1 but for $7^+$ yrast energies.\\
\end{tabular}

\bigskip
\section*{Graph 20. ~~~~~~~${\bold 9^+}$ yrast energies in even-even nuclei.}
\begin{tabular}{p{0.9cm}p{6in}}
&(Color online) Same as Graph 1 but for $9^+$ yrast energies.\\
\end{tabular}

\footnotesize 
\datatables

\newpage
\include{table1}

\newpage
\include{table2}

\newpage
\include{table3}

\newpage
\include{table4}

\normalsize


\newpage

\begin{Dfigures}[ht!]
\begin{center}
\includegraphics[width=0.75\linewidth]{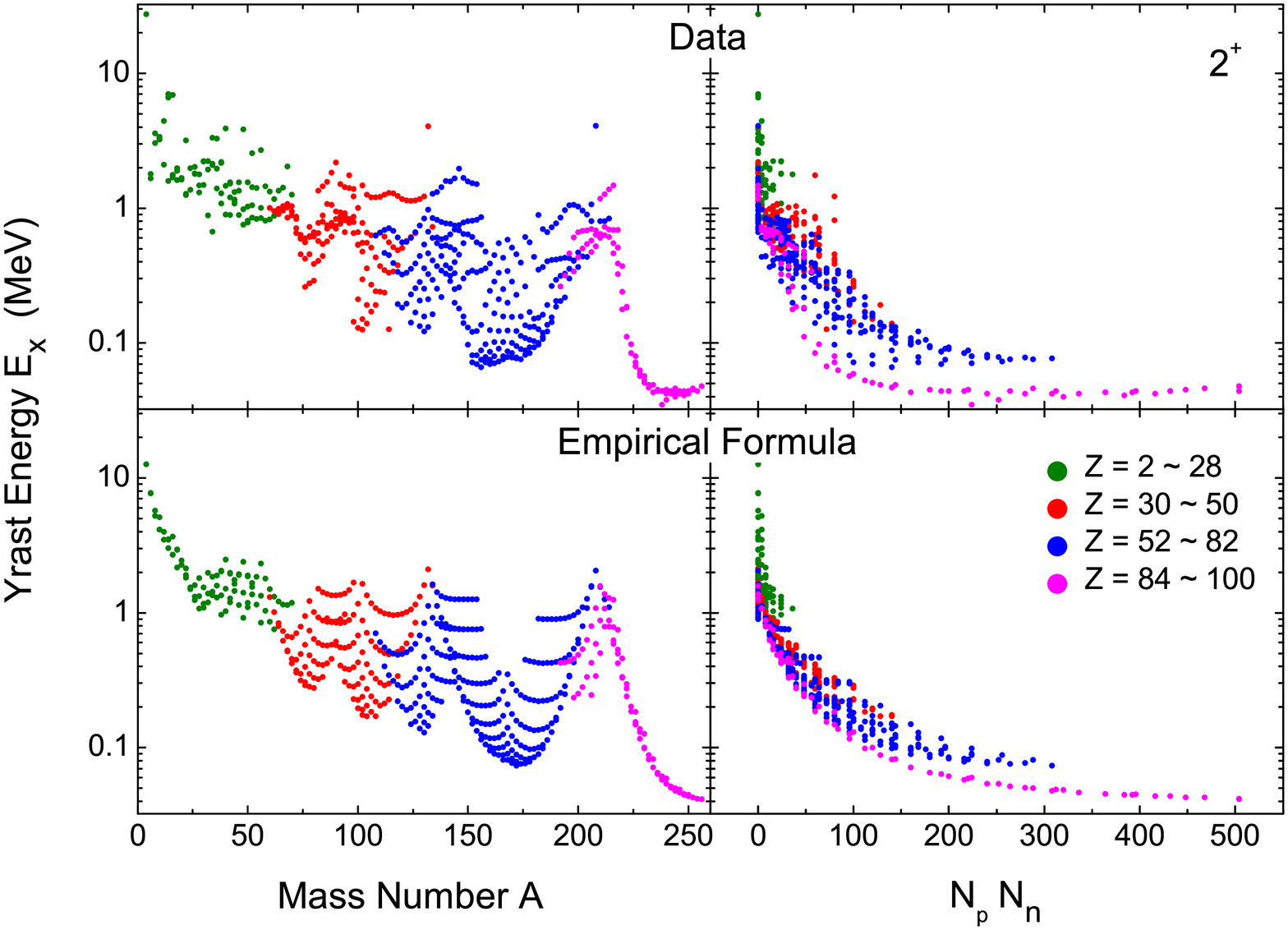}
\end{center}
\caption{$2^+$ yrast energies in even-even nuclei.}\label{graph-1}
\end{Dfigures}

\bigskip

\begin{Dfigures}[ht!]
\begin{center}
\includegraphics[width=0.75\linewidth]{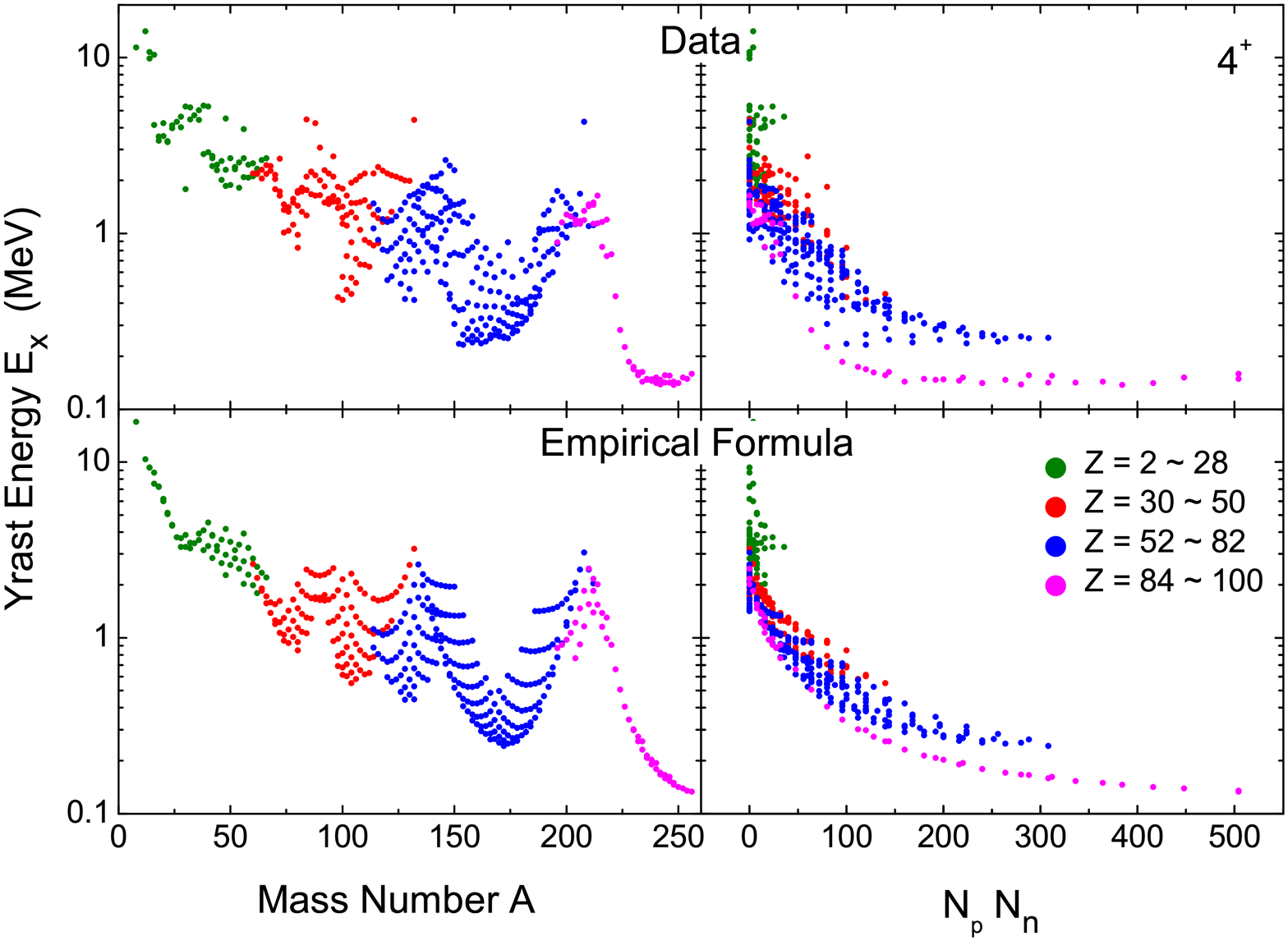}
\end{center}
\caption{$4^+$ yrast energies in even-even nuclei.}\label{graph-2}
\end{Dfigures}

\begin{Dfigures}[ht!]
\begin{center}
\includegraphics[width=0.75\linewidth]{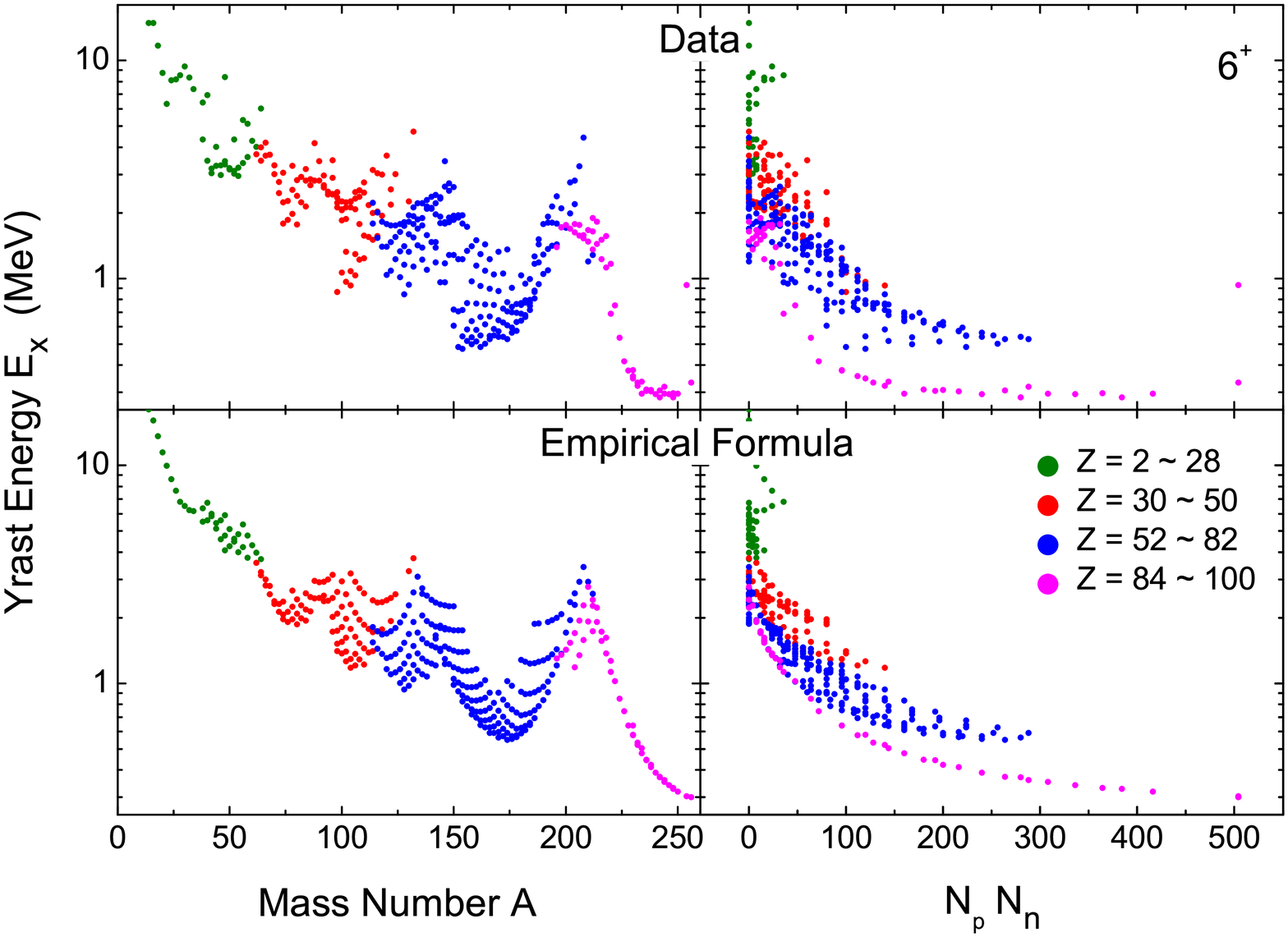}
\end{center}
\caption{$6^+$ yrast energies in even-even nuclei.}\label{graph-3}
\end{Dfigures}

\bigskip

\begin{Dfigures}[ht!]
\begin{center}
\includegraphics[width=0.75\linewidth]{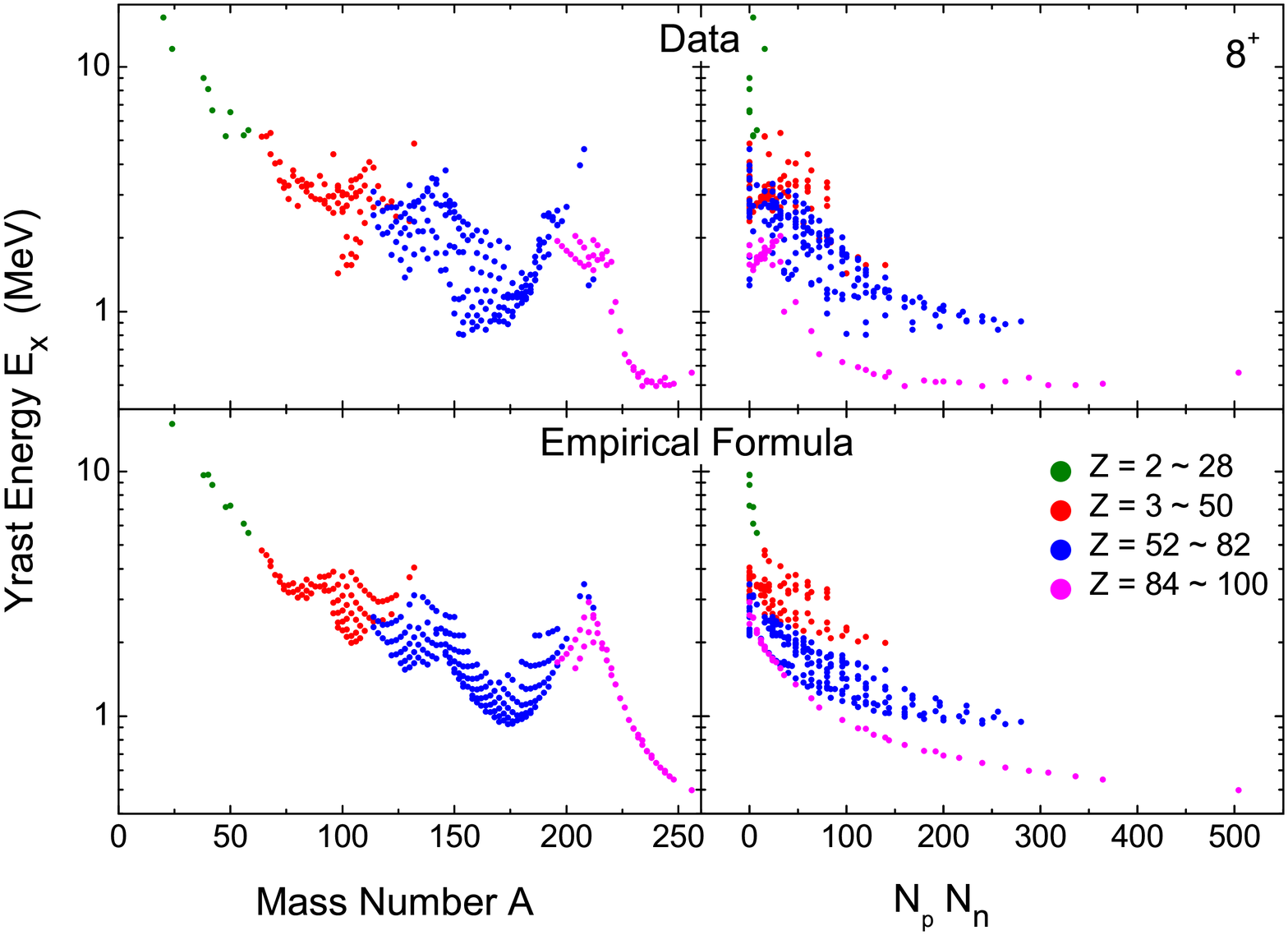}
\end{center}
\caption{$8^+$ yrast energies in even-even nuclei.}\label{graph-4}
\end{Dfigures}

\begin{Dfigures}[ht!]
\begin{center}
\includegraphics[width=0.75\linewidth]{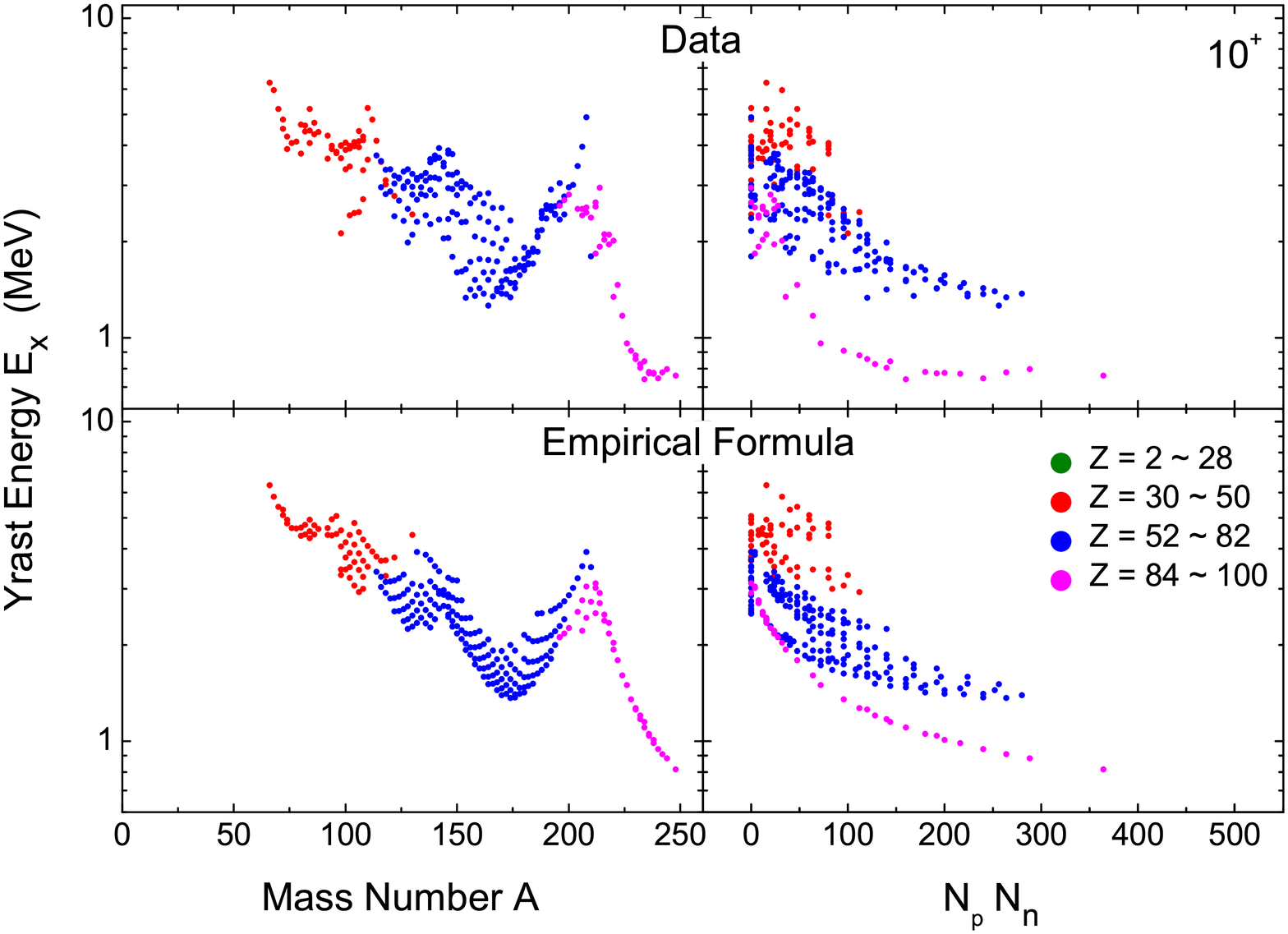}
\end{center}
\caption{$10^+$ yrast energies in even-even nuclei.}\label{graph-5}
\end{Dfigures}

\bigskip

\begin{Dfigures}[ht!]
\begin{center}
\includegraphics[width=0.75\linewidth]{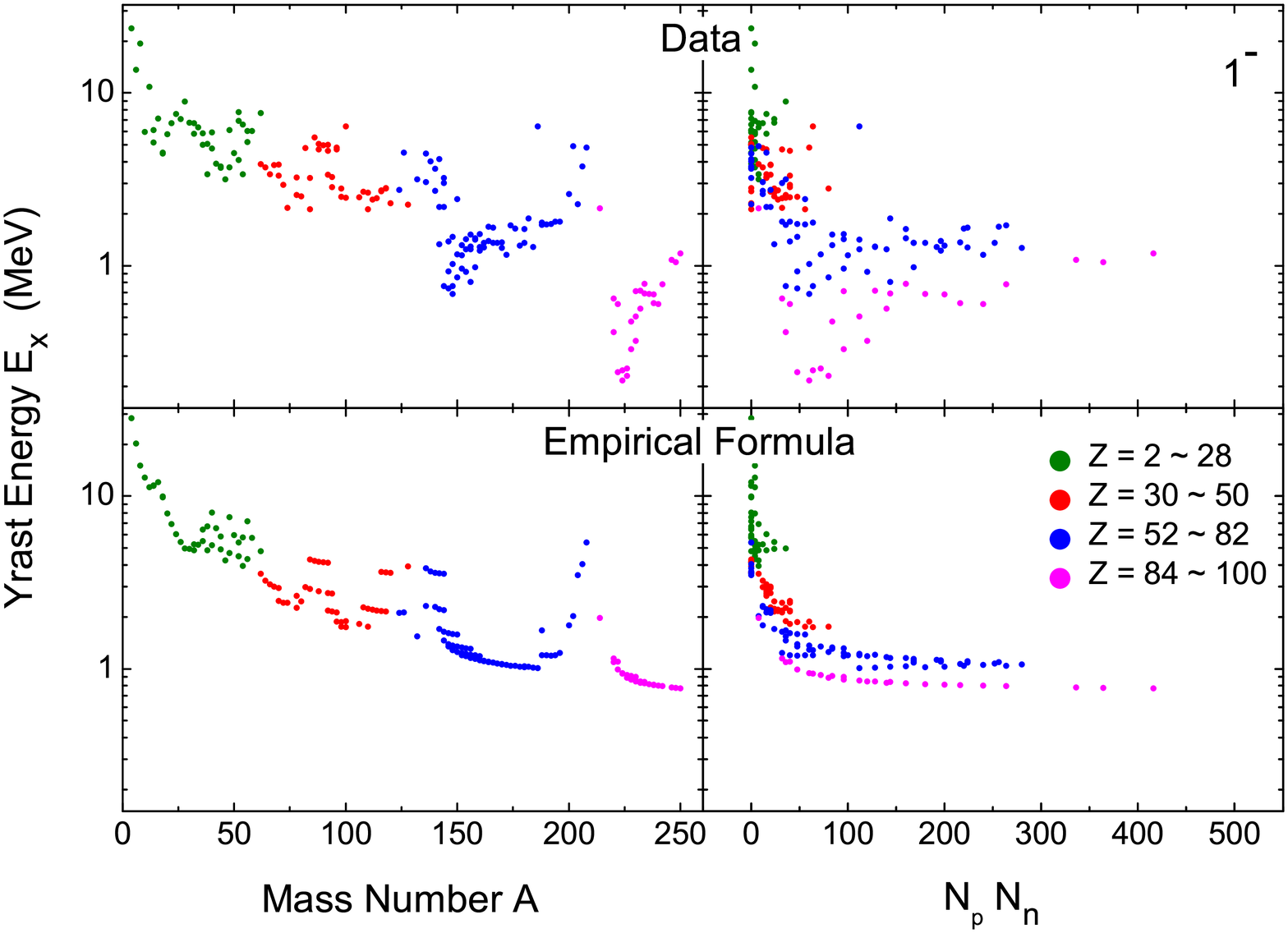}
\end{center}
\caption{$1^-$ yrast energies in even-even nuclei.}\label{graph-6}
\end{Dfigures}

\begin{Dfigures}[ht!]
\begin{center}
\includegraphics[width=0.75\linewidth]{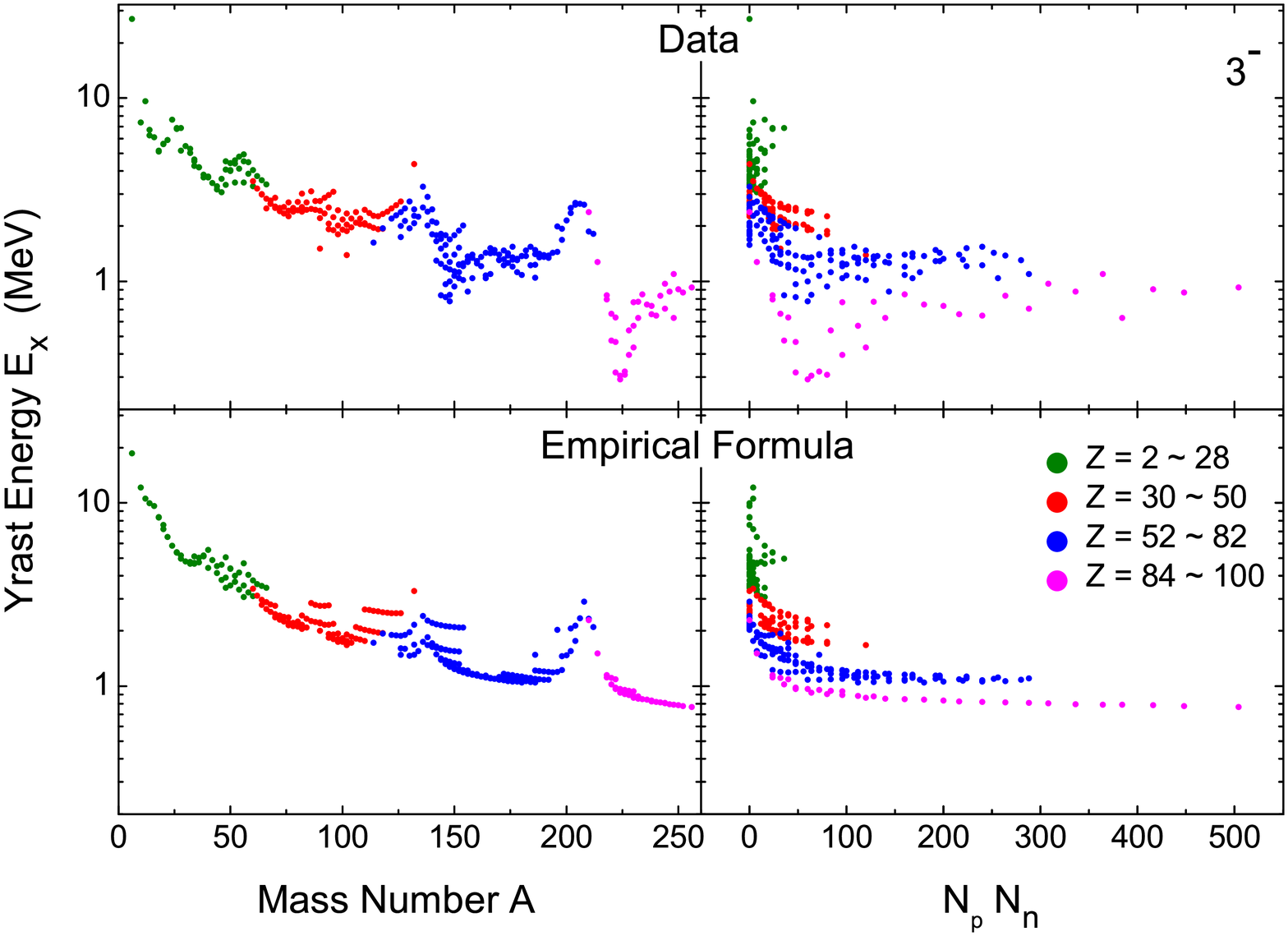}
\end{center}
\caption{$3^-$ yrast energies in even-even nuclei.}\label{graph-7}
\end{Dfigures}

\bigskip

\begin{Dfigures}[ht!]
\begin{center}
\includegraphics[width=0.75\linewidth]{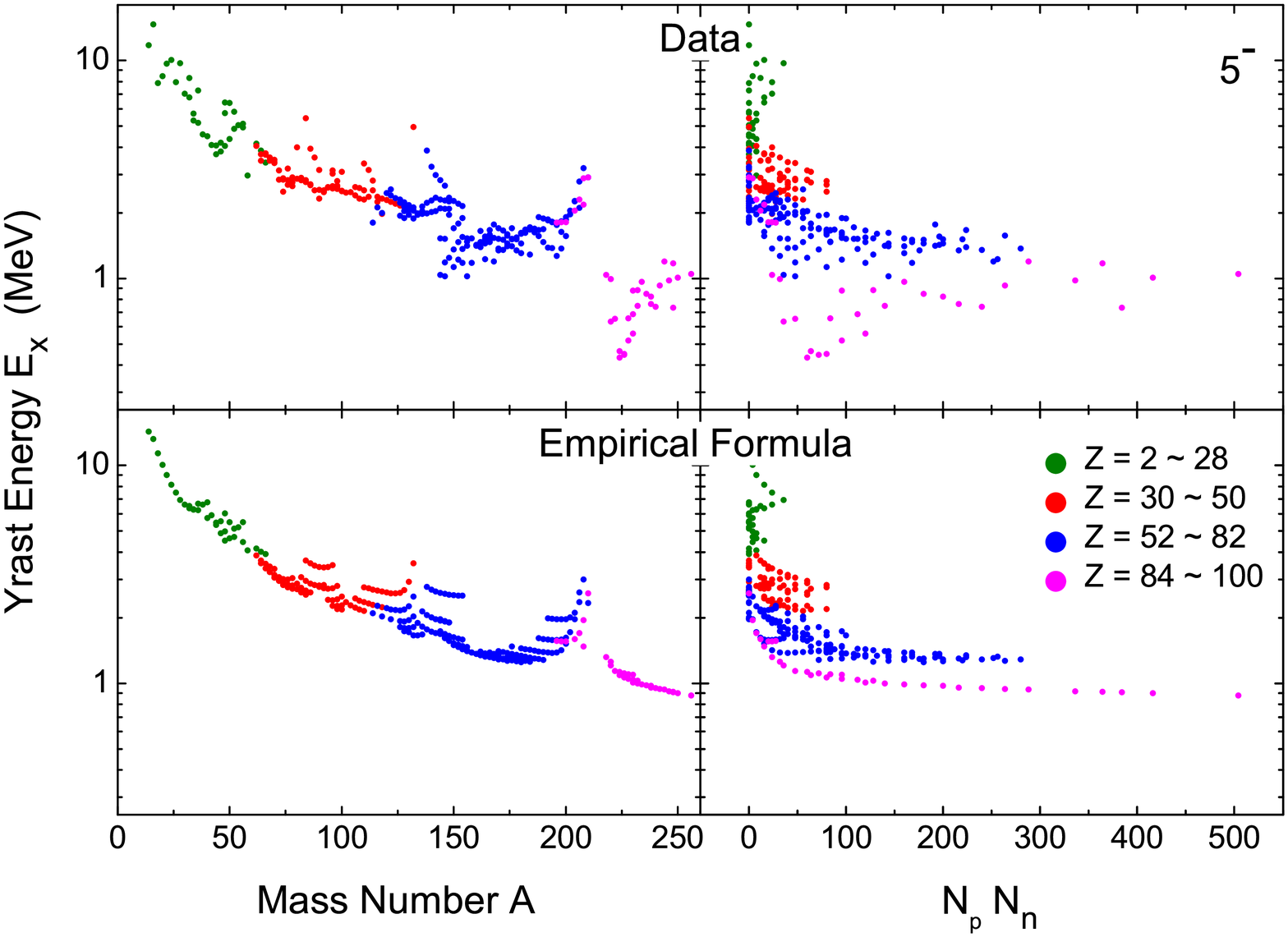}
\end{center}
\caption{$5^-$ yrast energies in even-even nuclei.}\label{graph-8}
\end{Dfigures}

\begin{Dfigures}[ht!]
\begin{center}
\includegraphics[width=0.75\linewidth]{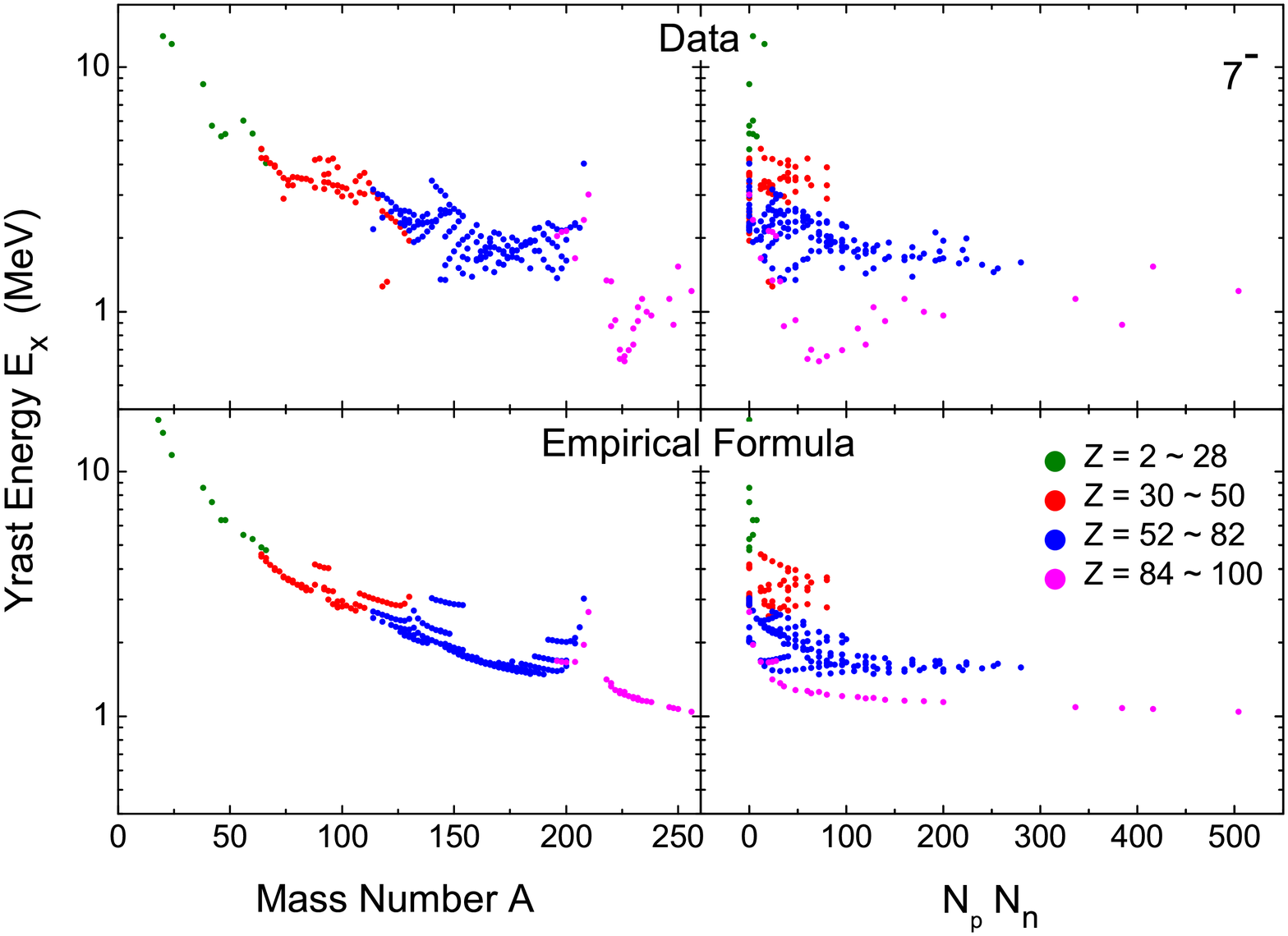}
\end{center}
\caption{$7^-$ yrast energies in even-even nuclei.}\label{graph-9}
\end{Dfigures}

\bigskip

\begin{Dfigures}[ht!]
\begin{center}
\includegraphics[width=0.75\linewidth]{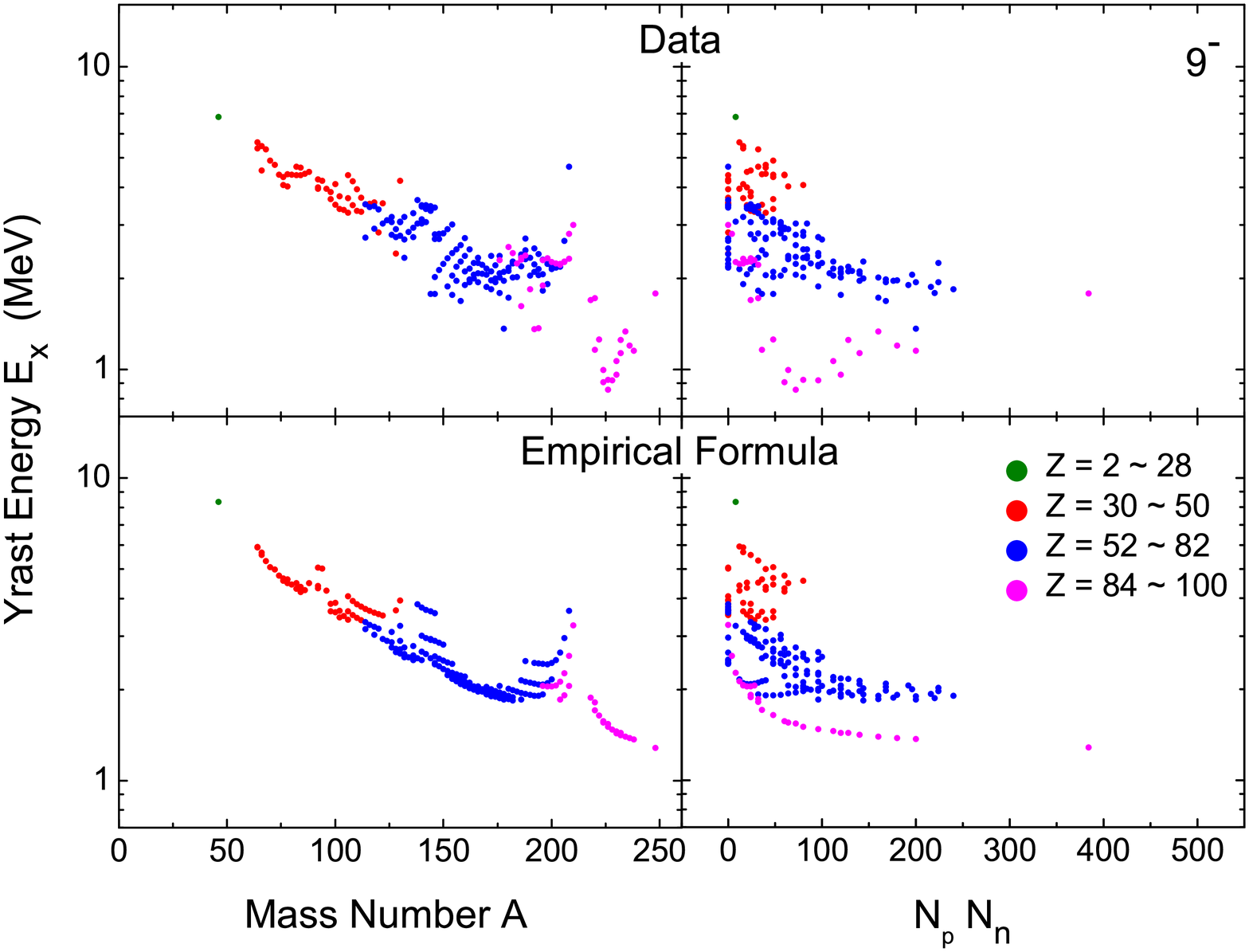}
\end{center}
\caption{$9^-$ yrast energies in even-even nuclei.}\label{graph-10}
\end{Dfigures}

\begin{Dfigures}[ht!]
\begin{center}
\includegraphics[width=0.75\linewidth]{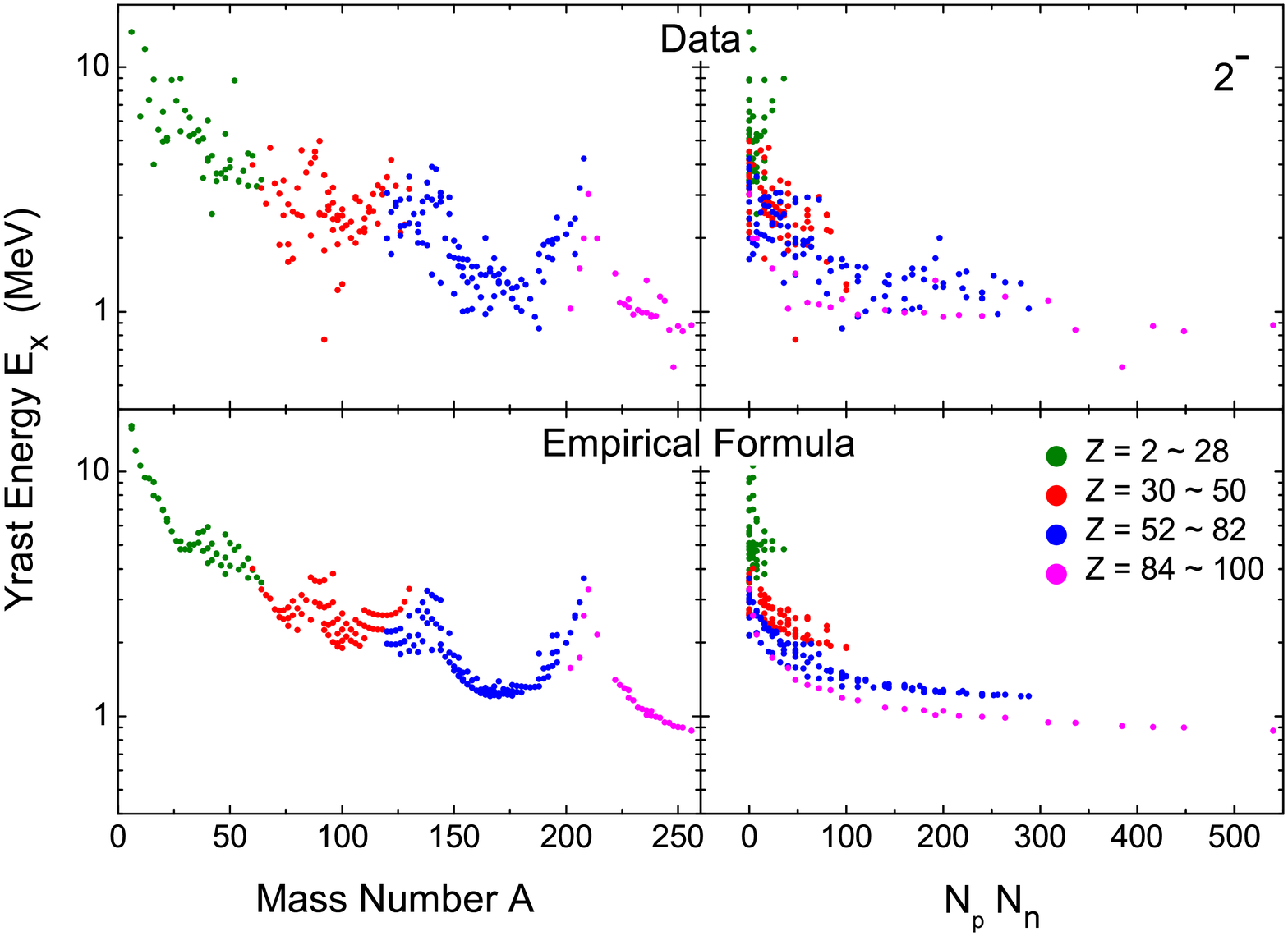}
\end{center}
\caption{$2^-$ yrast energies in even-even nuclei.}\label{graph-11}
\end{Dfigures}

\bigskip

\begin{Dfigures}[ht!]
\begin{center}
\includegraphics[width=0.75\linewidth]{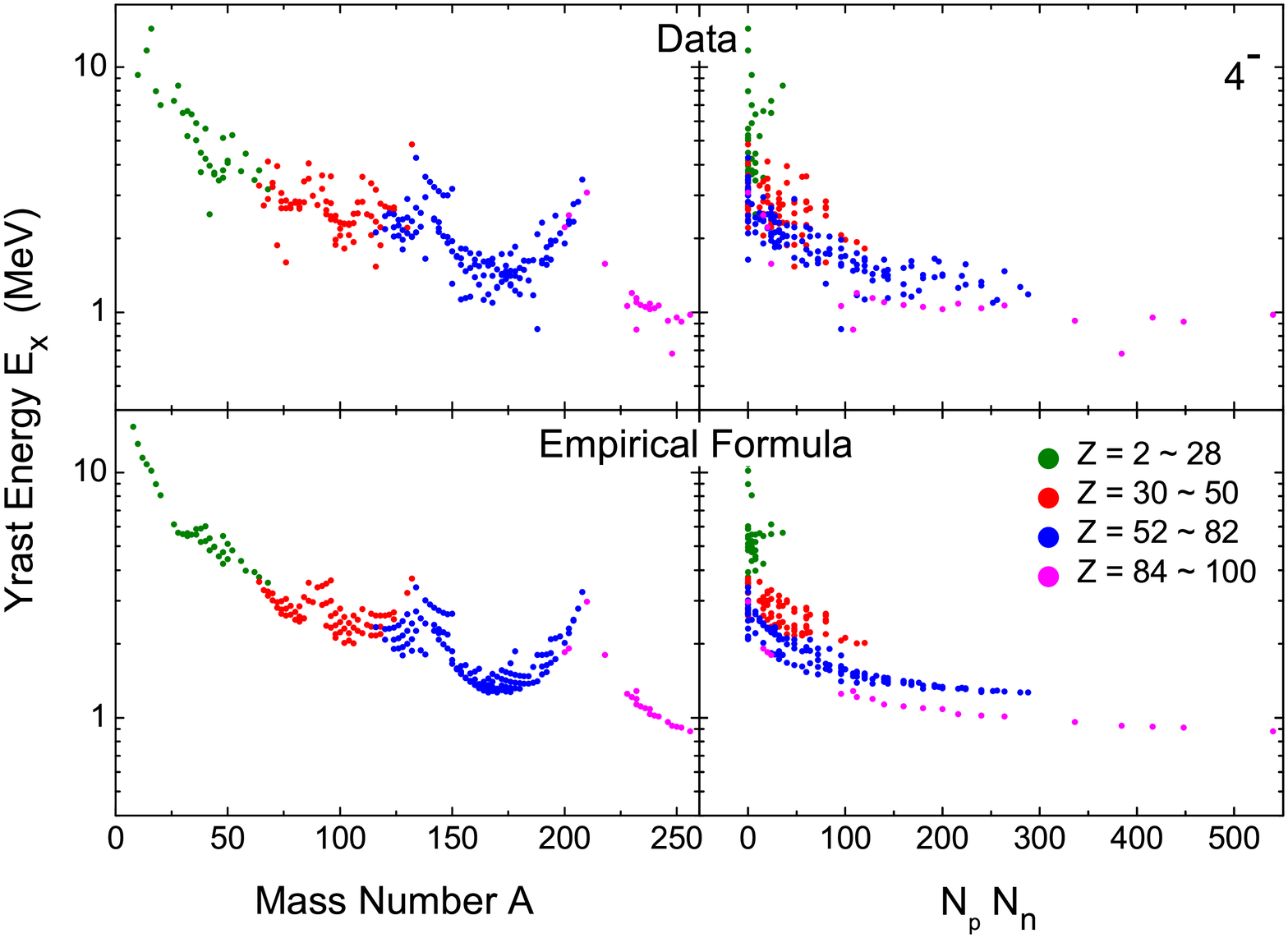}
\end{center}
\caption{$4^-$ yrast energies in even-even nuclei.}\label{graph-12}
\end{Dfigures}

\clearpage

\begin{Dfigures}[ht!]
\begin{center}
\includegraphics[width=0.75\linewidth]{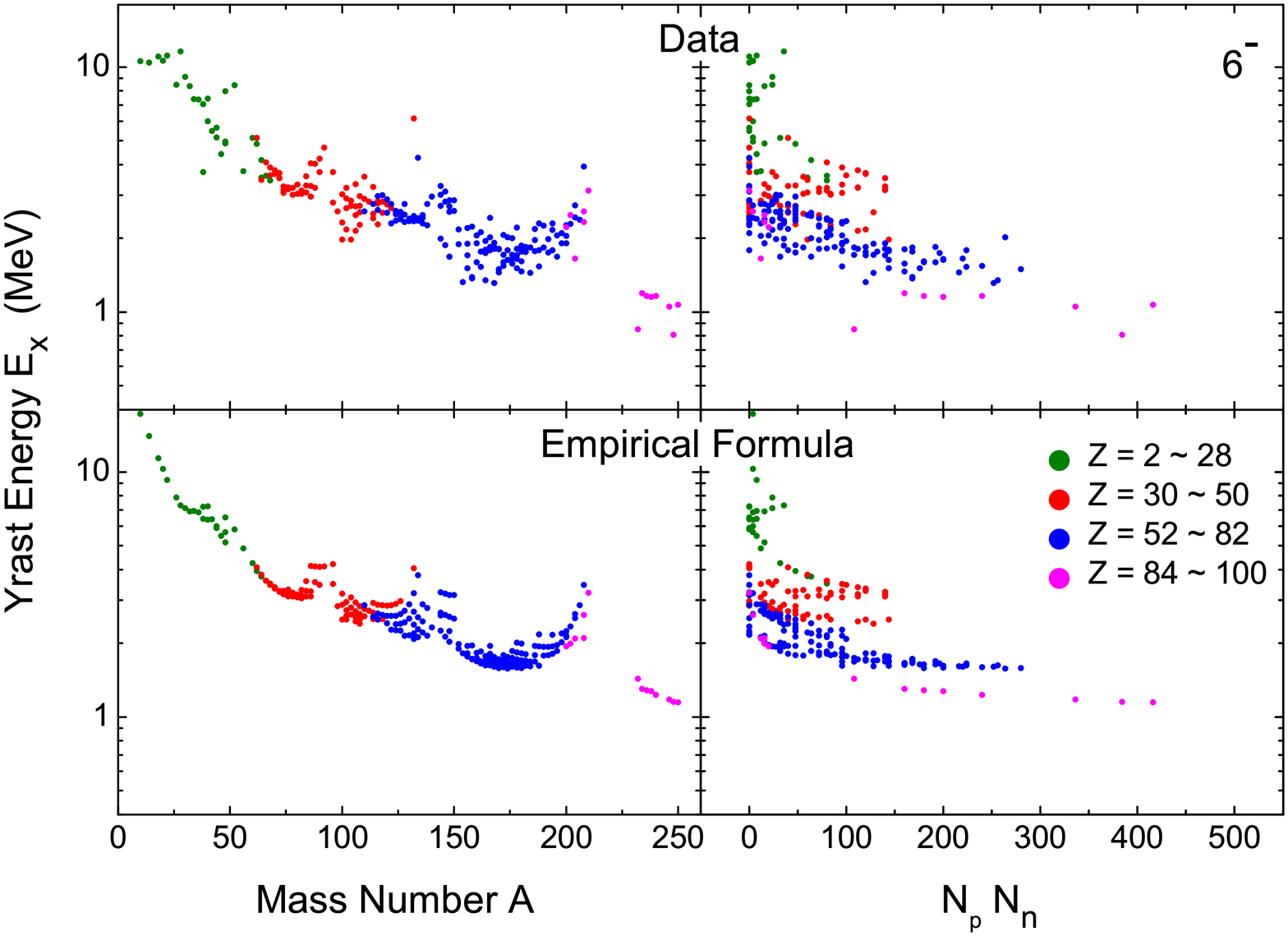}
\end{center}
\caption{$6^-$ yrast energies in even-even nuclei.}\label{graph-13}
\end{Dfigures}

\bigskip

\begin{Dfigures}[ht!]
\begin{center}
\includegraphics[width=0.75\linewidth]{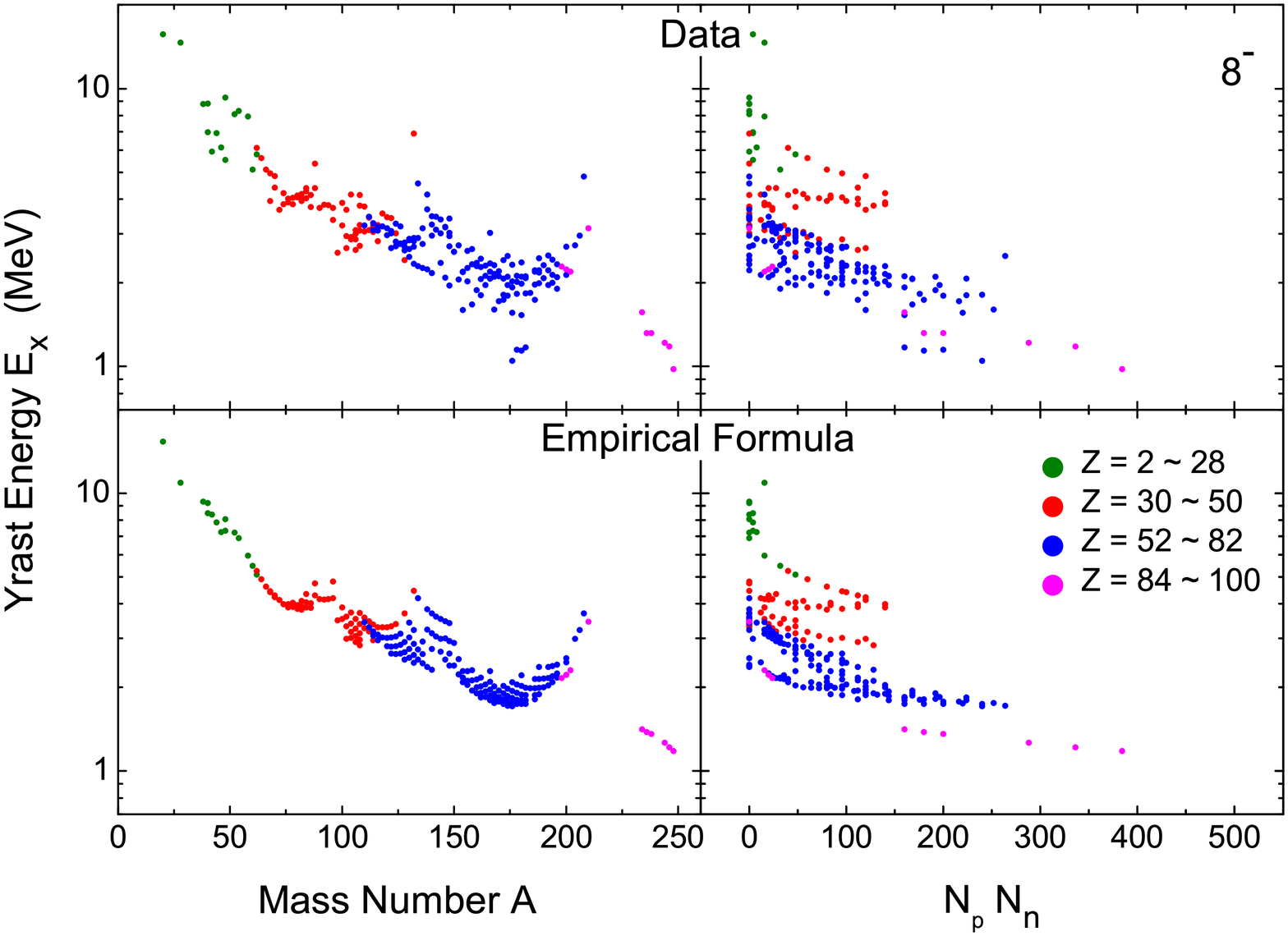}
\end{center}
\caption{$8^-$ yrast energies in even-even nuclei.}\label{graph-14}
\end{Dfigures}

\begin{Dfigures}[ht!]
\begin{center}
\includegraphics[width=0.75\linewidth]{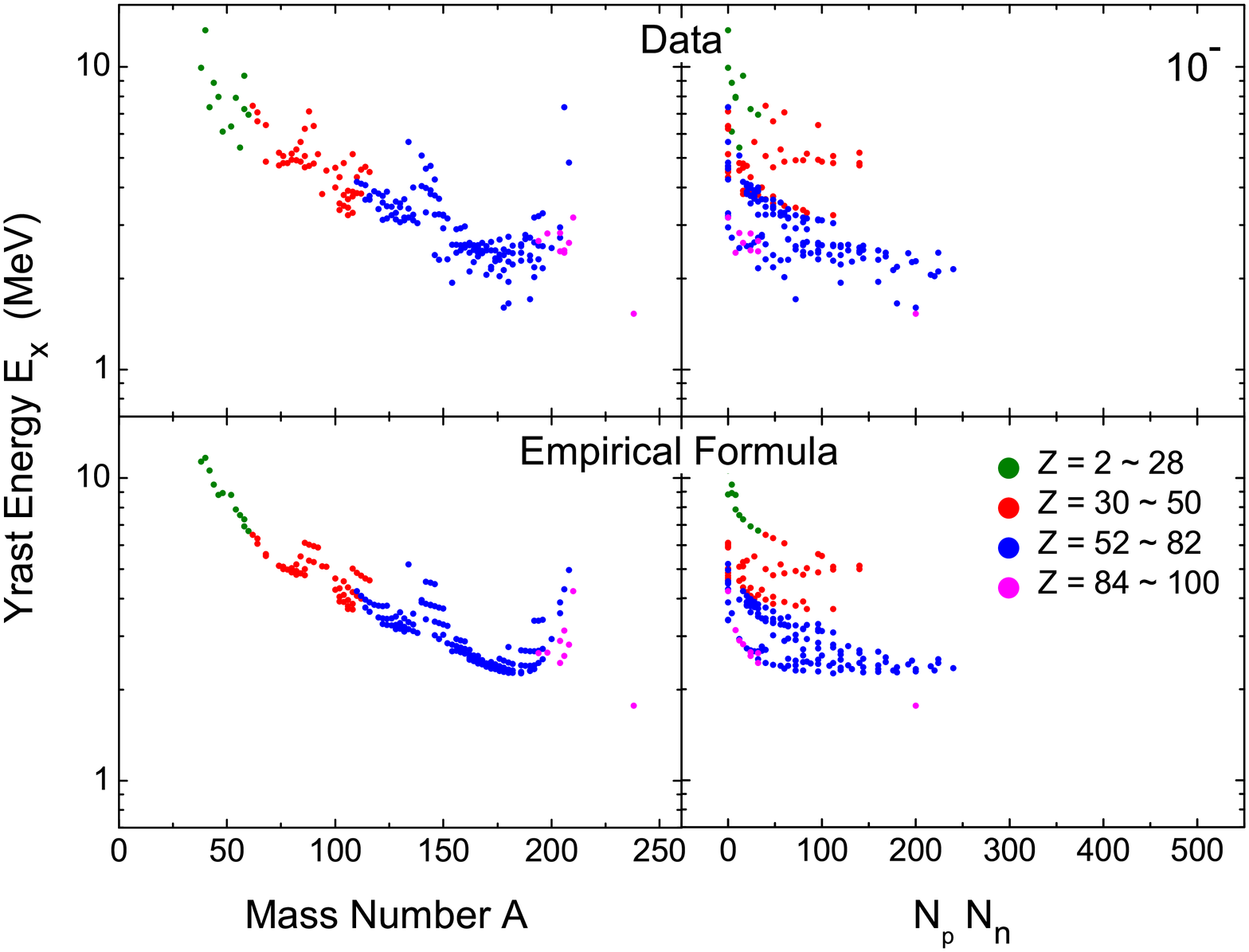}
\end{center}
\caption{$10^-$ yrast energies in even-even nuclei.}\label{graph-15}
\end{Dfigures}

\bigskip

\begin{Dfigures}[ht!]
\begin{center}
\includegraphics[width=0.75\linewidth]{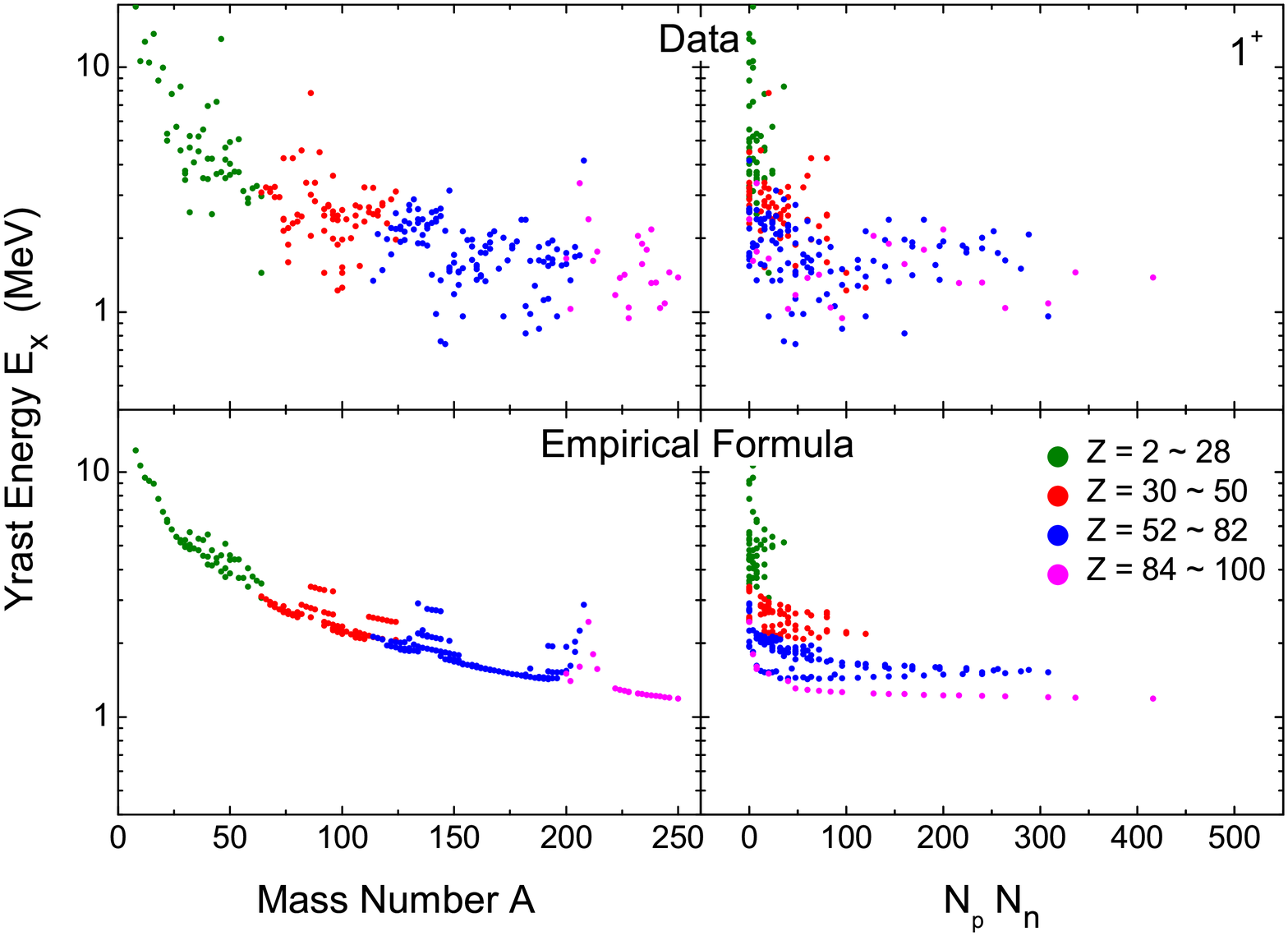}
\end{center}
\caption{$1^+$ yrast energies in even-even nuclei.}\label{graph-16}
\end{Dfigures}

\begin{Dfigures}[ht!]
\begin{center}
\includegraphics[width=0.75\linewidth]{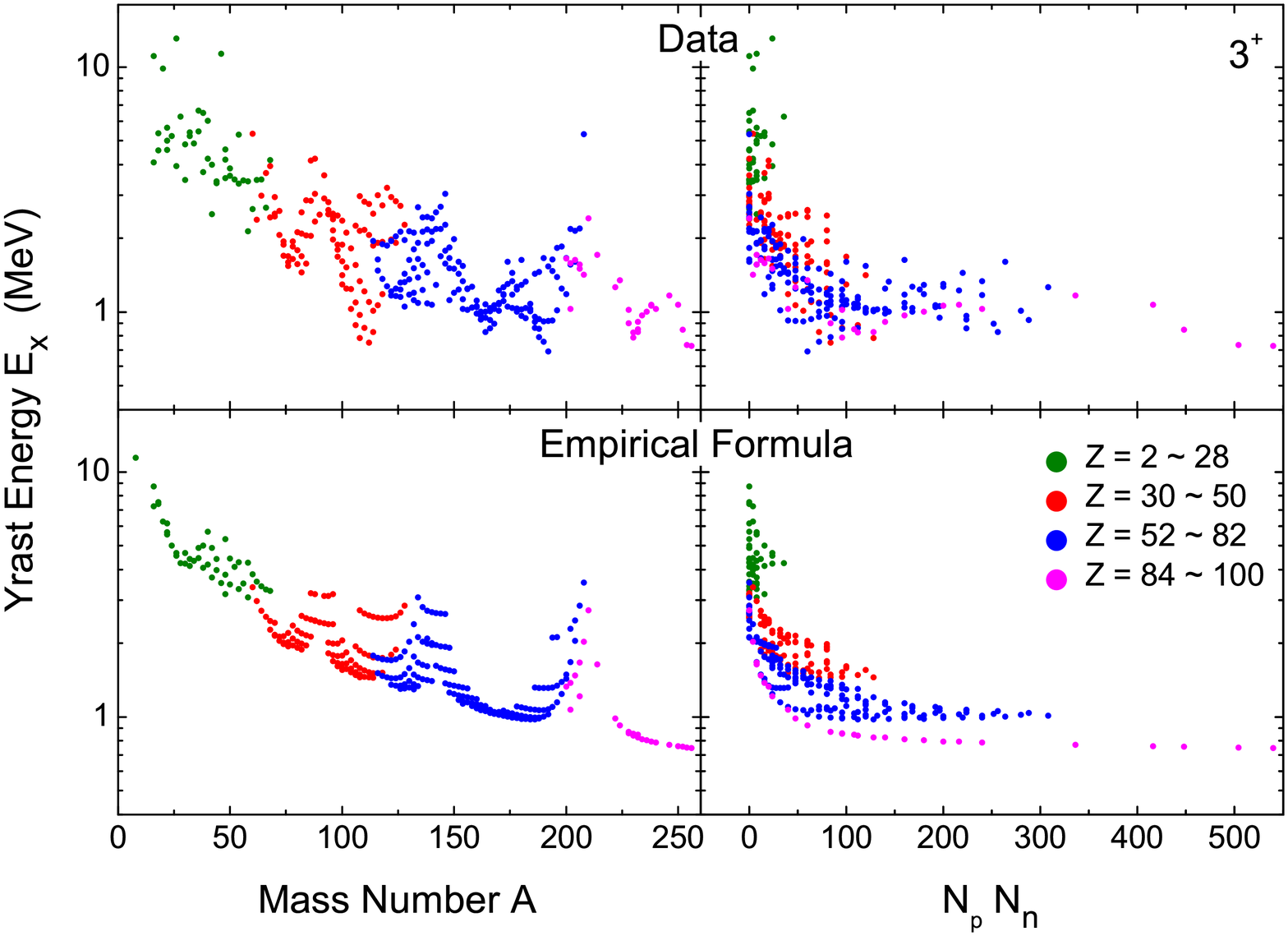}
\end{center}
\caption{$3^+$ yrast energies in even-even nuclei.}\label{graph-17}
\end{Dfigures}

\bigskip

\begin{Dfigures}[ht!]
\begin{center}
\includegraphics[width=0.75\linewidth]{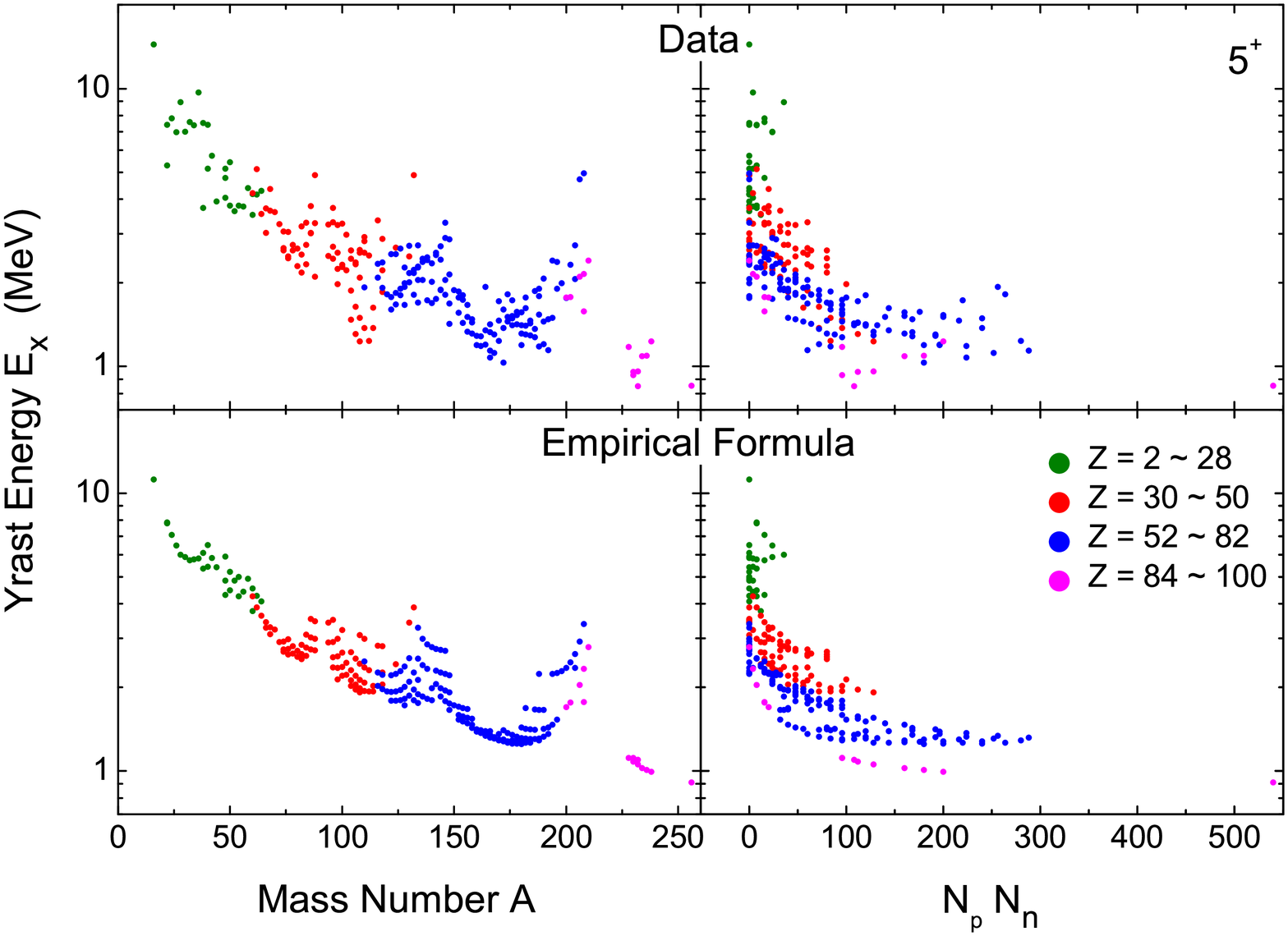}
\end{center}
\caption{$5^+$ yrast energies in even-even nuclei.}\label{graph-18}
\end{Dfigures}

\begin{Dfigures}[ht!]
\begin{center}
\includegraphics[width=0.75\linewidth]{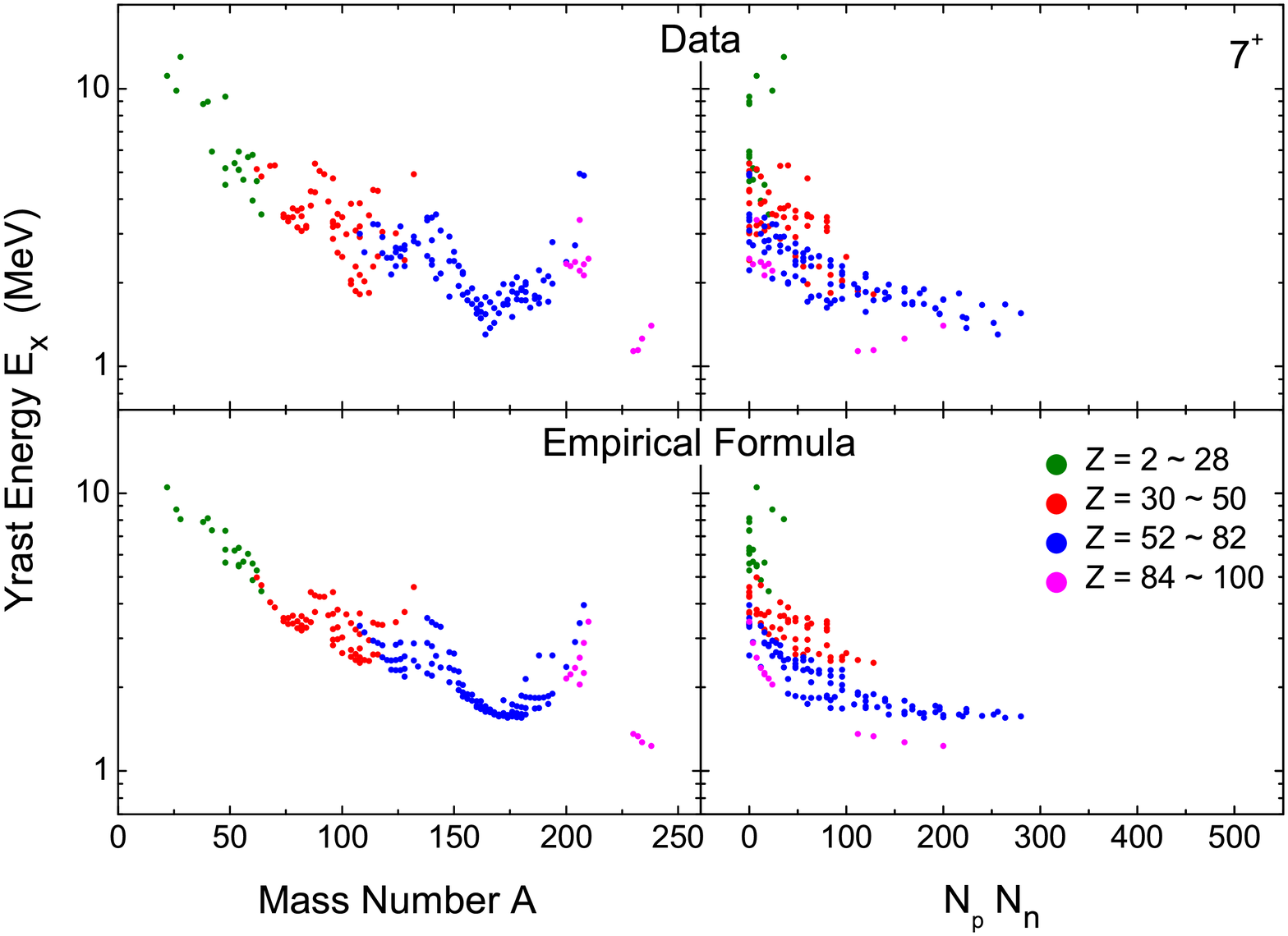}
\end{center}
\caption{$7^+$ yrast energies in even-even nuclei.}\label{graph-19}
\end{Dfigures}

\bigskip

\begin{Dfigures}[ht!]
\begin{center}
\includegraphics[width=0.75\linewidth]{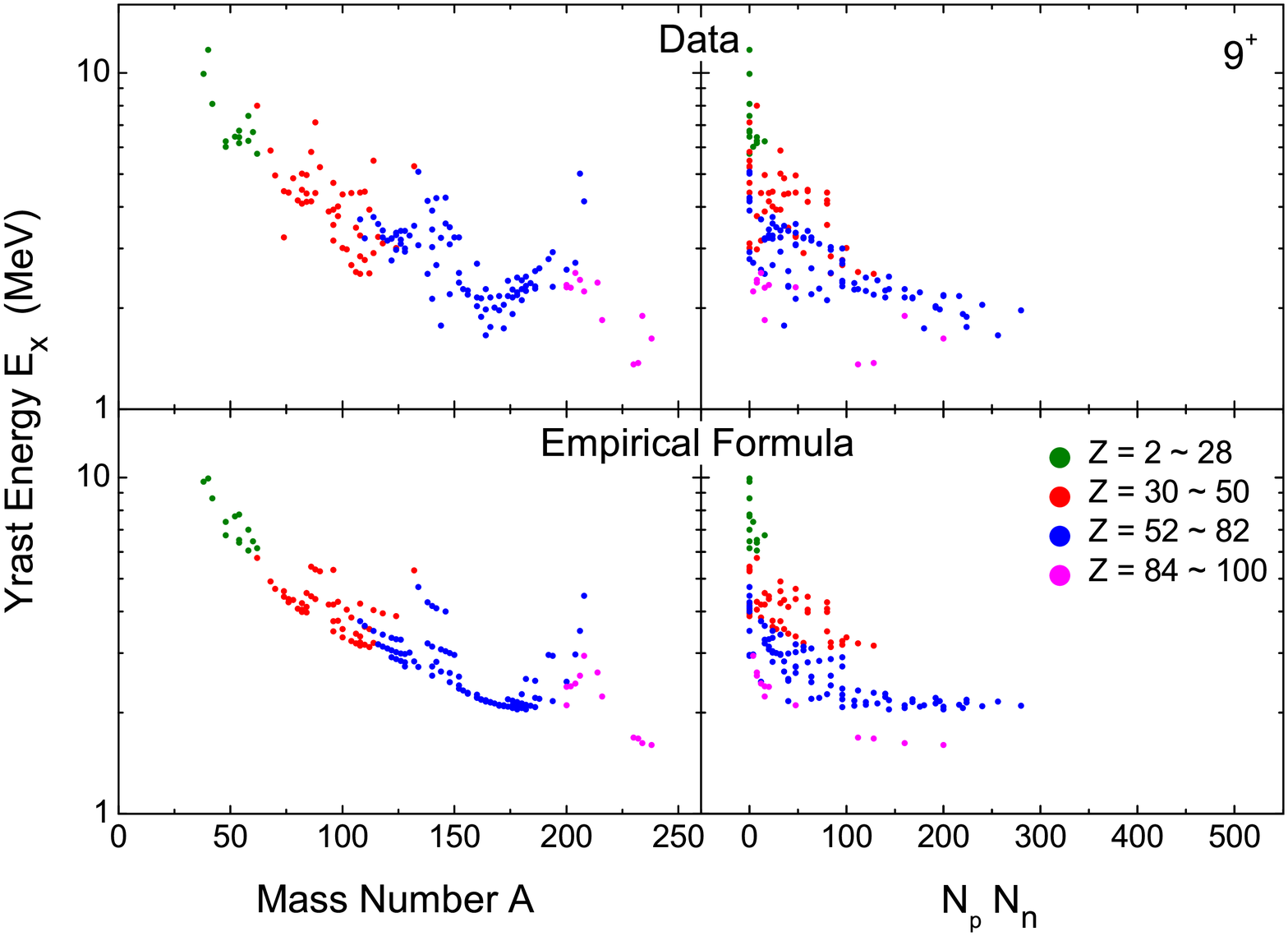}
\end{center}
\caption{$9^+$ yrast energies in even-even nuclei.}\label{graph-20}
\end{Dfigures}

\end{document}

%% file: table1.tex
\begin{center}
\begin{longtable}{@{\extracolsep\fill}ccccccccc@{}}

\caption{Yrast energies of the natural parity even multipole states
  in even-even nuclei.}\\
\hline
Nuclide&$A$&$N_p$&$N_n$&&&$E_x$~~(MeV)&&\\ \cline{5-9}
&&&&$~ J=2^{+}$&$~ J=4^{+}$&$~ J=6^{+}$&$~ J=8^{+}$&$~ J=10^{+}$\\
\hline
\endfirsthead
\multicolumn{9}{l}{(Table I. continued)}\\
\hline
Nuclide&$A$&$N_p$&$N_n$&&&$E_x$~~(MeV)&&\\ \cline{5-9}
&&&&$~ J=2^{+}$&$~ J=4^{+}$&$~ J=6^{+}$&$~ J=8^{+}$&$~ J=10^{+}$\\
\hline
\endhead
\multicolumn{9}{r}%
{(continued on next page)}
\endfoot

\hline
\endlastfoot											
He&4&0&0&27.42/{\textcolor{blue}{12.67}}&&&&\\
&6&0&2&1.797/{\textcolor{blue}{7.695}}&&&&\\
&8&0&2&3.590/{\textcolor{blue}{5.713}}&&&&\\
&10&0&0&3.240/{\textcolor{blue}{5.125}}&&&&\\
&&&&&&&&\\
Be&6&2&0&1.670/{\textcolor{blue}{7.725}}&&&&\\
&8&2&2&3.040/{\textcolor{blue}{5.241}}&11.40/{\textcolor{blue}{16.89}}&&&\\
&10&2&2&3.368/{\textcolor{blue}{4.152}}&&&&\\
\end{longtable}
\end{center}

%% file: table2.tex
\begin{center}
\begin{longtable}{@{\extracolsep\fill}ccccccccc@{}}

\caption{Yrast energies of the natural parity odd multipole states
  in even-even nuclei.}\\
\hline
Nuclide&$A$&$N_p$&$N_n$&&&$E_x$~~(MeV)&&\\ \cline{5-9}
&&&&$~ J=1^{-}$&$~ J=3^{-}$&$~ J=5^{-}$&$~ J=7^{-}$&$~ J=9^{-}$\\
\hline
\endfirsthead
\multicolumn{9}{l}{(Table 2. continued)}\\
\hline
Nuclide&$A$&$N_p$&$N_n$&&&$E_x$~~(MeV)&&\\ \cline{5-9}
&&&&$~ J=1^{-}$&$~ J=3^{-}$&$~ J=5^{-}$&$~ J=7^{-}$&$~ J=9^{-}$\\
\hline
\endhead
\multicolumn{9}{r}%
{(continued on next page)}
\endfoot

\hline
\endlastfoot											
He&4&0&0&23.64/{\textcolor{blue}{28.34}}&&&&\\
&6&0&2&13.60/{\textcolor{blue}{20.17}}&&&&\\
&&&&&&&&\\
Be&6&2&0&&27.00/{\textcolor{blue}{12.15}}&&&\\
&8&2&2&19.40/{\textcolor{blue}{15.07}}&&&&\\
&10&2&2&5.960/{\textcolor{blue}{12.80}}&7.371/{\textcolor{blue}{10.56}}&&&\\
&&&&&&&&\\
C&12&2&2&10.84/{\textcolor{blue}{11.24}}&9.641/{\textcolor{blue}{9.939}}&&&\\
\end{longtable}
\end{center}

%% file: table3.tex
\begin{center}
\begin{longtable}{@{\extracolsep\fill}ccccccccc@{}}

\caption{Yrast energies of the unnatural parity even multipole states
  in even-even nuclei.}\\
\hline
Nuclide&$A$&$N_p$&$N_n$&&&$E_x$~~(MeV)&&\\ \cline{5-9}
&&&&$~ J=2^{-}$&$~ J=4^{-}$&$~ J=6^{-}$&$~ J=8^{-}$&$~ J=10^{-}$\\
\hline
\endfirsthead
\multicolumn{9}{l}{(Table 3. continued)}\\
\hline
Nuclide&$A$&$N_p$&$N_n$&&&$E_x$~~(MeV)&&\\ \cline{5-9}
&&&&$~ J=2^{-}$&$~ J=4^{-}$&$~ J=6^{-}$&$~ J=8^{-}$&$~ J=10^{-}$\\
\hline
\endhead
\multicolumn{9}{r}%
{(continued on next page)}
\endfoot

\hline
\endlastfoot

He&4&0&0&21.84/{\textcolor{blue}{20.20}}&&&&\\
&6&0&2&13.90/{\textcolor{blue}{14.97}}&&&&\\
&&&&&&&&\\
Be&6&2&0&26.00/{\textcolor{blue}{15.36}}&23.00/{\textcolor{blue}{19.50}}&&&\\
&8&2&2&18.91/{\textcolor{blue}{12.15}}&20.90/{\textcolor{blue}{15.37}}&&&\\
&10&2&2&6.263/{\textcolor{blue}{10.56}}&9.270/{\textcolor{blue}{13.07}}&10.57/{\textcolor{blue}{17.30}}&&\\
&&&&&&&&\\
C&12&2&2&11.83/{\textcolor{blue}{9.444}}&19.55/{\textcolor{blue}{11.48}}&&&\\
\end{longtable}
\end{center}

%% file: table4.tex
\begin{center}
\begin{longtable}{@{\extracolsep\fill}ccccccccc@{}}

\caption{Yrast energies of the unnatural parity odd multipole states
  in even-even nuclei.}\\
\hline
Nuclide&$A$&$N_p$&$N_n$&&&$E_x$~~(MeV)&&\\ \cline{5-9}
&&&&$~ J=1^{+}$&$~ J=3^{+}$&$~ J=5^{+}$&$~ J=7^{+}$&$~ J=9^{+}$\\
\hline
\endfirsthead
\multicolumn{9}{l}{(Table 4. continued)}\\
\hline
Nuclide&$A$&$N_p$&$N_n$&&&$E_x$~~(MeV)&&\\ \cline{5-9}
&&&&$~ J=1^{+}$&$~ J=3^{+}$&$~ J=5^{+}$&$~ J=7^{+}$&$~ J=9^{+}$\\
\hline
\endhead
\multicolumn{9}{r}%
{(continued on next page)}
\endfoot

\hline
\endlastfoot

He&4&0&0&28.31/{\textcolor{blue}{20.22}}&&&&\\
&&&&&&&&\\
Be&8&2&2&17.64/{\textcolor{blue}{12.26}}&19.07/{\textcolor{blue}{11.42}}&&&\\
&10&2&2&10.57/{\textcolor{blue}{10.63}}&&&&\\
&&&&&&&&\\
C&12&2&2&12.71/{\textcolor{blue}{9.468}}&&&&\\
&14&2&0&10.45/{\textcolor{blue}{9.214}}&&&&\\
&16&2&2&&4.088/{\textcolor{blue}{7.231}}&&&\\
\end{longtable}
\end{center}